\documentclass
[12pt,tightenlines,eqsecnum,floats,aps,amsmath,amssymb,tightenlines,nofootinbib,prd,shownopacs,notitlepage]{revtex4}
\usepackage{graphicx,verbatim}
\usepackage{amsmath}
\usepackage{amsfonts}
\usepackage{amssymb}
\usepackage[colorinlistoftodos]{todonotes}
\usepackage{epstopdf}

\def\t{\tilde}

\def\psym{\Gamma_{\rm YM}}
\def\psymc{\mathbb{C}\Gamma_{\rm YM}}
\def\psymr{\Gamma_{\rm YM}^{\rm Riem}}
\def\psyml{\Gamma_{\rm YM}^{\rm Lor}}
\def\psgeo{\Gamma_{\rm geo}}

\def\f{\frac}
\def\Lie{\mathcal{L}}
\def\P{\tilde{p}}

\usepackage{psfrag}
\usepackage{accents}
\numberwithin{equation}{section}
\def\be{\begin{equation}}
\def\ee{\end{equation}}
\def\ba{\begin{eqnarray}}
\def\ea{\end{eqnarray}}
\def\bi{\begin{itemize}}
\def\ei{\end{itemize}}

\def\qz{\mathring{q}}
\def\ez{\mathring{\epsilon}}
\def\V{\tilde{\mathcal{V}}}
\def\Sv{\vec{S}}
\def\Nv{\vec{N}}
\def\Mv{\vec{M}}
\def\G{\mathcal{G}}
\def\S{\tilde{\tilde{\mathcal{S}}}}
\def\rmd{{\rm d}}

\def\l{\mathcal{L}}
\def\gcl{\mathfrak{L}}
\def\ggcl{\mathbb{L}}
\def\vins{\mathbin{\dimen0=\ht\strutbox  \divide\dimen0 by 2
 \hbox{\vbox{\hrule width\dimen0}\hskip-0.4pt\vrule height\dimen0}}\,}
\def\mfd{\mathfrak{D}}

\def\E{\tilde E}
\def\Eub{\tilde{\underbar{E}}}
\def\N{\underaccent{\tilde}{N}}
\def\M{\underaccent{\tilde}{M}}
\def\D{\mathcal{D}}
\def\kub{\underbar{k}}

\begin{document}

\title{Gravitational dynamics: \\A novel shift in the Hamiltonian paradigm}

\author{Abhay Ashtekar${}^{1}\,$}
\email{ashtekar.gravity@gmail.com} 
\author{Madhavan Varadarajan${}^{2}\,$}
\email{madhavan@rri.res.in}
\affiliation{
${}^{1}$ Institute for Gravitation and the Cosmos \& Physics Department, The Pennsylvania State University, University Park, PA 16802 U.S.A.\\
${}^{2}$  Raman Research Institute, Bangalore-560 080, India
}

\begin{abstract}
It is well known that Einstein's equations assume a simple polynomial  form in the Hamiltonian framework based on a Yang-Mills phase space \cite{newvar}. We re-examine the gravitational dynamics in this framework and show that {\em time} evolution of the gravitational field can be re-expressed as (a gauge covariant generalization of) the Lie derivative along a novel shift vector field in {\em spatial} directions.  Thus, the canonical transformation generated by the Hamiltonian constraint acquires a geometrical interpretation on the Yang-Mills phase space, similar to that generated by the diffeomorphism constraint.  In classical general relativity this geometrical interpretation significantly simplifies calculations and also illuminates the relation between dynamics in the `integrable' (anti)self-dual sector \cite{newman,plebanski,penrose} and in the full theory.  For quantum gravity, it provides a point of departure to complete the Dirac quantization program for general relativity in a  more satisfactory fashion \cite{mv1}. This gauge theory perspective may also be helpful in extending the `double copy' ideas \cite{bernetal} relating the Einstein and Yang-Mills dynamics to a non-perturbative regime.  Finally, the notion of generalized, gauge covariant Lie derivative may also be of interest to the mathematical physics community as it hints at some potentially rich structures that have not been explored.

\end{abstract}
\maketitle

\section{Introduction} 
\label{s1}

The conceptual framework of general relativity (GR) has provided us with the richest example of the interplay between geometry and physics to date.  On the other hand, the mathematical structure of Einstein's equations is complicated:  they constitute a set of non-linear partial differential equations that are highly non-polynomial in the basic variable --the space-time metric. It is desirable both from the point of view of classical general relativity, as well as for progress in non-perturbative quantum gravity,  to deepen our structural understanding of these equations. In this paper, we pursue this goal in the context of the Hamiltonian formulation of general relativity. As is well-known, in this formulation Einstein's equations split into two sets: the gravitational constraints which are conditions on the initial data on a Cauchy slice, and the evolution equations which describe how the data evolve off the Cauchy slice with respect to any given choice of time.  Our new results pertain to the evolution equations.

In most of this paper, we will restrict ourselves to vacuum GR where issues of central interest to our discussion reside. We use a parameter $\epsilon$ that takes the value 1 while referring to the Riemannian signature and -1 while referring to the Lorentzian signature. Finally, \emph{a tilde will denote a field with density weight 1 and an under-tilde, of -1.} (However, as a concession to notational simplicity, we make an exception and omit the tildes on the square-root of the determinant of the metric which has density weight $1$, and its inverse which has density weight $-1$.)

Let us begin by briefly recalling the geometrodynamical phase space framework, introduced by Arnowitt, Deser and Misner (ADM)  (see, e.g., \cite{adm}). The  canonically conjugate variables consist of a positive definite metric $q_{ab}$ on a spatial 3-manifold $\Sigma$ that serves as the configuration variable, and its conjugate momentum is a symmetric tensor density $\P^{ab}$ of weight $1$ on $\Sigma$, both in the Lorentzian (-,+,+,+) and Riemannian (+,+,+,+)  signature: 
\be \label{pbadm} \{ q_{ab}(x),\, \P^{cd}(y) \} \,=\, \delta_{(a}^{c}\, \delta_{b)}^{d}\, \delta(x,y) \ee 
When equations of motion hold, $\P^{ab}$ encodes the extrinsic curvature $k_{ab}$ of $\Sigma$ via $\P^{ab} \!=  -\epsilon\, q^{\f{1}{2}} (k^{ab} - k q^{ab})$ where $q^{\f{1}{2}}$ denotes the square root of the determinant of $q_{ab}$ and $k = q^{cd}k_{cd}$.  % 
The phase space variables $(q_{ab}, \P^{ab})$ are subject to a set of  4 first class constraints:
\be \label{admcon}
\t{C}^{a} := -2 D_{a}\, \P^{ab} \simeq 0  \quad {\rm and} \quad  {\t{C} :=  - \,q^{\f{1}{2}} \,\mathcal{R} - \epsilon\, q^{-\f{1}{2}}\, \big(q_{ac} q_{bd} - {\textstyle{\f{1}{2}}}\, q_{ab} q_{cd} \big)\,  \P^{ab}\, \P^{cd}\,  \simeq \, 0\, ,} \ee
where $D$ is the torsion-free derivative operator compatible with $q_{ab}$; and $\mathcal{R}$, the scalar curvature of the metric $q_{ab}$. In the standard terminology they are, respectively, the diffeomorphism and the Hamiltonian constraints. As in any generally covariant  theory, they are intertwined with dynamics: the Hamiltonian generating evolution equations is a 
linear combination of constraints. The action of the diffeomorphism constraint $\t{C}^{a}$ is simple:  When smeared with a shift vector field $S^{a}$, it generates the infinitesimal canonical transformations
\be \label{admdiffeo}
\dot{q}_{ab} =  \Lie_{\vec{S}}\, q_{ab} \quad {\rm and} \quad \dot{\P}^{ab} =  \Lie_{\vec{S}}\, \P^{ab}\, 
\ee
where, as usual, $\Lie_{\vec{S}}$ denotes the Lie derivative with respect to $S^{a}$.   By contrast, because of the presence of the determinant factors and the Ricci scalar $\mathcal{R}$ in the Hamiltonian constraint, the evolution equations it generates are highly non-polynomial in the  fundamental canonical variables $(q_{ab}, \P^{ab})$, and do not admit a natural geometric interpretation (see Eq. \ref{admevo2}). Our goal is to reformulate the Hamiltonian framework in such a way that time evolution becomes  simple, with a transparent  geometric interpretation analogous to that of (\ref{admdiffeo}). 

A natural starting point towards this goal is to seek alternate canonical variables in terms of which the constraint and evolution equations simplify.  Such a reformulation has been available 
in the literature for some quite time now:  the \emph{connection-dynamics}  framework \cite{newvar,aabook}.  It served two purposes. First,  the constraint and evolution equations simplified enormously in that they became low order polynomials in the fundamental canonical variables. Second, GR was brought closer to gauge theories that describe the other three fundamental interactions.  Specifically, gravitational issues could be investigated using the conceptual setting and well developed techniques of gauge theories. 

Let us then start ab-initio, introduce a background independent  $SU(2)$  gauge theory without any reference to gravity, and summarize its relation to geometrodynamics at the end.  Thus, now the configuration variable is a connection 1-form $A_{a}^{i}$  on a 3-manifold $\Sigma$  that takes values in  the Lie-algebra $su(2)$ of $SU(2)$, (so that the index $i$ refers to $su(2)$).  Its conjugate momentum is an $su(2)$-valued electric field $\E^{a}_{i}$ on  $\Sigma$ (with density weight $1$, since there is no background metric).   Although the phase space is the same as $\psym$ of the $SU(2)$ Yang-Mills (YM) theory in Minkowski space, there is a key difference in the formulation of dynamics since the space-time metric is not given a priori but is rather the end-product of dynamics. Thus, in stark contrast to the YM theory,  the constraints and evolution equations that $A_{a}^{i}$ and $\E^{a}_{i}$ are now subject to, cannot involve a pre-specified space-time metric.  What are the simplest gauge covariant equations one can construct from this canonical pair $(A_{a}^{i},\, \E^{a}_{i})$ without reference to any background field? These are:
\be \label{constraints} \t\G_{i} := \D_{a} \E^{a}_{i}  \simeq 0; \quad  {\V}_{a} :=  \E^{b}_{i} F_{ab}^{i} \simeq 0;\quad {\rm and} \quad  {\S} := \ez^{ij}{}_{k}\E^{a}_{i} \E^{b}_{j} F_{ab}^{k} \simeq 0 \, ; \ee
where $\D$ is the gauge covariant derivative operator defined by $A_{a}^{i}$; \, $F_{ab}^{k}$ its curvature; and $\ez^{ij}{}_{k}$ the structure constants of $SU(2)$. %(Note that although we do not have a metric $q_{ab}$ on $\Sigma$,  the \emph{internal} indices can be lowered and raised using the Cartan-Killing metric $\qz_{ij}$ on the Lie algebra $su(2)$.) 
The first equation is at most linear in each of the canonical variables; the second is linear in $\E^{a}_{i}$ and at most quadratic in $A_{a}^{i}$; and the third is at most quadratic in each variable. Thus all equations are low order polynomials. Since, by construction, they do not involve any time derivatives, they serve as constraint equations on the gauge theory phase space. Note that the first equation in (\ref{constraints})  is the familiar Gauss constraint on $\psym$ that generates local gauge transformations. The second and third are new from the gauge theory perspective. Interestingly, on $\psym$ the full set turns out to be of first class  in the Dirac classification. Since we have 9 configuration variables $A_{a}^{i}$ per point of $\Sigma$ and 7 constraints, we have the kinematics of a \emph{background independent} gauge theory with \emph{precisely 2 degrees of freedom.} One may therefore hope that this is just a reformulation of GR in a gauge theory disguise.%Interestingly, if one sets up an appropriate dictionary, Eqs.  (\ref{constraints})  turn out provide us \emph{precisely} with constraint equations of general relativity.%
\footnote{The next simplest gauge invariant term is cubic in $E^{a}_{i}$ and leads to GR with a cosmological constant.}

This expectation is correct.  The detailed dictionary between the gauge theory and geometrodynamics is given in Section \ref{s2}.  The correspondence enables us to recast the second and the third  constraints in (\ref{constraints}) in the language of geometrodynamics:  Once the Gauss constraint is satisfied, they turn out to be precisely the diffeomorphism and Hamiltonian constraints (\ref{admcon})!  Thus, if we think of the gauge theory phase space as primary, then the constraints of GR are the simplest low order polynomials in the basic canonical variables that can be written down without reference to a background field. Since evolution equations are obtained by taking the Poisson brackets of  the canonical variables with constraints, they also contain only low order polynomials in the new canonical pair.  These features brought out the fact that gravitational dynamics simplifies greatly if one adopts a gauge theory perspective in analyzing Einstein's theory \cite{newvar,aabook,cdj,cdjm,alrev,ttbook,30years}. Therefore the connection-dynamics formulation lies at the  foundation of Loop Quantum Gravity (LQG).

There is an additional feature of this reformulation that was also noted in the early days but has not been directly used. Let us consider the fundamental, spin $\f{1}{2}$ representation of  $SU(2)$ to express the constraints. Then the second and the third of the constraints in (\ref{constraints}) naturally blend together into a single equation (see, e.g. \cite{aabook}, page 85)
\be \label{blending1}
\E^{a}_{A}{}^{B}\,  \E^{b}_{B}{}^{C} \, F_{ab\, C}{}^{D}  \simeq 0 \ee
where the indices $A,\, B$ refer to $SU(2)$ spinors. The trace-free part of this unified constraint is precisely the second of Eqs. (\ref{constraints}) --(essentially) the diffeomorphism constraint-- and the trace-free part to the third -- (essentially) the Hamiltonian constraint. Thus the framework naturally combines  the diffeomorphism and the Hamiltonian constraints of GR into a single, simple expression.%
\footnote{This fact was not further exploited because, at that time, the emphasis was on further simplifying the Poisson algebra of constraints. Although the fusion (\ref{blending1})  is elegant,  it did not further simplify the algebra.}
This observation suggests that it should be possible to blend the canonical transformation generated by the two constraints as well. In particular, although the intuition shaped by the ADM framework has led us to consider  `time-evolution'  to be on a \emph{very different} footing from  `space-evolution' (\ref{admdiffeo}),  the fusion (\ref{blending1}) of their generators suggests that the `time' evolution should also admit a natural geometric interpretation analogous to the `space-evolution' (\ref{admdiffeo}). 

 In this paper we will show that this expectation is borne out in detail in the following sense: \emph{Time} evolution will emerge as a gauge covariant generalization of the Lie derivative of the canonical variables, along  a {\em spatial}  shift vector field $N^{a}_{i}:= \N \E^{a}_{i}$ constructed from the lapse function $\N$.  This  novel shift  is a `q-number' vector field in the sense that it depends on the momentum variable, the YM electric field $\E^{a}_{i}$.  We will therefore refer to it as the \emph{electric shift}.  This reformulation of the action of the Hamiltonian constraint provides new conceptual insights already at the classical level. Furthermore,  analysis of certain model systems that mimic GR  \cite{2+1u13,3+1u13,mvlong} indicates that these novel shift vector fields are likely to play a role in constructing a more satisfactory Hamiltonian constraint operator in LQG.  Taken together, these considerations suggest a new avenue to the problem of obtaining an anomaly-free realization of the constraint algebra in quantum theory that faithfully mirrors the structure of its classical counterpart \cite{mv1}. 

The paper is organized as follows. In Section \ref{s2} we fix the notation and recall how geometrodynamics emerges from the YM phase space.  The section contains the basic equations from the connection formulation of GR that will be used in the rest of the paper.  In Section \ref{s3} we introduce the notion of a \emph{generalized gauge covariant Lie derivative}, now associated with vector fields that also carry $su(2)$ internal indices, such as the electric shift $N^{a}_{i}$. We use it in Section \ref{s4} to bring out a new geometrical structure that underlies the dynamics of GR.  Our central result recasts `time evolution'  in terms of `space-evolution' along electric shifts $N^{a}_{i}$ in both Riemannian and Lorentzian signatures. The existing literature in Hamiltonian LQG generally follows the original derivation  of connection variables  \cite{newvar} and discusses the canonical transformations generated by constraints using  ordinary Lie derivatives. Then the actions of the diffeomorphism and Hamiltonian constraints appear quite distinct. The use of generalized gauge covariant Lie derivatives along electric shifts brings out the conceptual unity underlying the two actions. Consequently, the derivation of the Poisson algebra of constraints simplifies significantly.  This formulation also shows the sense in which dynamics of  full GR is a natural, gauge invariant  generalization of the much simpler equations governing dynamics of (anti)self-dual, `integrable' sector. In Section \ref{s5}, we first summarize the main results and then put them in a broader context.  Appendix A provides further details on generalized, gauge covariant Lie derivatives, using the algebraic and operational approach to differential geometry due to Penrose \cite{rp,rpwr} and Geroch \cite{rg} that puts to the forefront properties of differential operations that are used repeatedly in practice. 

The interplay between YM theory and GR has experienced a renaissance recently, thanks to the 
astute observation that the perturbative scattering amplitudes of classical GR can be expressed using ``double copy'' relations between the gauge theory and gravitational amplitudes (for a review, see \cite{bernetal}). Because of the focus on on-shell quantities, manifest gauge covariance plays an important role in that analysis. At a perturbative level, satisfaction of the Jacobi identity by the so-called kinematical factors that appear in the expression of scattering amplitudes implies that the gravitational amplitudes are invariant under linearized diffeomorphisms.  The reformulation of \emph{non-perturbative} Einstein dynamics of this paper may help clarify the `origin' of this connection, since YM kinematics and strong emphasis on gauge covariance play a central role in our analysis. On the gravitational phenomenology side,  the double copy framework provides new approximation methods for analyzing gravitational waves in binary coalescence of compact objects (see, e.g. \cite{doublecopy}).  In particular, these methods have been used to calculate the Hamiltonians used in the Post-Newtonian approximation. Because ours is a Hamiltonian approach, it could provide a more direct route.  Therefore, to improve accessibility to the double copy community, we have made a special effort to make the presentation reasonably self-contained at the cost of repeating some discussion that the LQG community is quite familiar with.  Also, since the  compactness and simplicity of time-evolution in the final result is quite surprising even from the LQG perspective, we present the intermediate steps in detail.

\section{From the YM phase space to Geometrodynamics}
\label{s2}

The primary focus of this paper is  on time-evolution of gravitational fields in 4-dimensional spacetimes $(M, g_{ab})$ of GR with  topology $M=\Sigma \times R$.  We will assume that the 3-manifold $\Sigma$ is oriented. {For simplicity of presentation, in the main text we will also assume that $\Sigma$ is compact without boundary.} In the asymptotically flat case, the Hamiltonians generating `space' and `time' evolution acquire surface terms in addition to the constraints, but this difference is not relevant for the main point under discussion in this paper. We will see in Section \ref{s4.3} that the principal results hold also in that case.  {Finally, we will mostly restrict ourselves to vacuum GR but briefly comment on inclusion of a scalar field as the source in Section \ref{s4.3}.}

In this section we fix notation and sketch the steps that enable one to regard GR as the simplest background independent gauge theory. Although the metric and the associated Riemannian geometry is invaluable in the classical domain, from this new perspective they are completely absent to begin with, and can be regarded as secondary notions that can be introduced at the end. The LQG community is very familiar with the technical arguments that lead to this paradigm shift \cite{newvar,cdj,cdjm}. However, one often starts from geometrodynamics and then passes to the gauge theory framework  (see, e.g., \cite{aabook,alrev,ttbook} and Chapter 1 in\cite{30years}). Consequently, the Riemannian structures used in geometrodynamics are at the forefront  in the beginning, and at the back of one's mind even when one arrives to the gauge theory framework. Here we adopt the opposite viewpoint and start with the $SU(2)$ YM phase space, without \emph{any} reference to a spacetime metric or gravity.  While this reversal is technically rather straightforward,  the concepts and mathematical tools it puts at the forefront suggest new directions that, in turn, bring out new structures.  As explained in Section \ref{s1}  the final result is an unforeseen perspective on time evolution in GR .

Let us then start with an $SU(2)$ gauge theory.  Since $SU(2)$ bundles over  3-manifolds $\Sigma$ are trivial, they admit global cross-sections which are naturally isomorphic with $\Sigma$. Therefore, any $SU(2)$ connection $\D$ can be regarded as acting on fields $\lambda_{i \ldots j}{}^{k \ldots l}$ on $\Sigma$ whose indices take values in $su(2)$, i.e.,  have only internal indices. We will restrict ourselves to those $\D$ that are compatible with the Cartan-Killing metric $\qz_{ij}$ on $su(2)$ and the structure constants $\ez_{ij}{}^{k}$ of $su(2)$, both regarded as fields on $\Sigma$:
\be  \D_{a} \qz_{ij} =0\qquad {\rm and} \qquad  \D_{a} \ez_{ij}{}^{k} =0\, . \ee
Consequently one can freely raise and lower internal indices of fields before or after the derivative operator $\D$ acts on them.  As usual, curvature of $\D$ will be denoted by $F_{ab}{}^{j}$:
 \be \label{curv} 2\D_{[a} \D_{b]} \lambda_{i} =: \ez_{ij}{}^{k}\, F_{ab}{}^{j} \,\lambda_{k} \, .\ee
Note that, although we will describe Einstein dynamics, this formulation \emph{does not} need  a metric tensor $q_{ab}$ on $\Sigma$, or the derivative operator $D$ compatible with it, while both are omnipresent in equations of geometrodynamics. Indeed, one does not need \emph{any} specific derivative operator on tensor fields.  In intermediate steps of calculations, it is often convenient to extend the action of $\D$ to tensor indices. However, throughout our analysis one can extend it using \emph{any torsion-free derivative operator.} It is generally simplest --although by no means necessary-- to extend the action of $\D$ using a flat  derivative operator on tensor indices and denote this extension by $\partial$. We will generally adopt this extension in the intermediate calculations.

Fix any flat connection --i.e. with zero curvature both on tensor and internal indices--  and denote it also by $\partial$.  Then since bundles under consideration are trivial, we can express the action of $\D$ via a globally defined connection 1-form $A_{a}^{i}$ on $\Sigma$:
\be  \label{A} \D_{a} \lambda_{i} =    \partial_{a} \lambda_{i} + \ez_{ij}{}^{k} A_{a}^{j}\lambda_{k} \
\quad \hbox{\rm so that} \quad F_{ab}^{i} = 2\partial_{[a} A_{b]}^{i} + \ez^{\,i}{}_{jk}\, A_{a}^{j} A_{b}^{k}\, .\ee

As explained in Section \ref{s1}, in the gauge theory formulation, the configuration variable is an $su(2)$ connection 1-form $A_{a}^{i}$ on $\Sigma$ and  its canonical momentum is the electric field $\E^a_i$,  a vector density of weight $1$ that also takes values in $su(2)$. Thus, the Poisson brackets are:
\be \label{fpb1} \{A_{a}^{i} (x),\, \E^{b}_{j} (y) \}  = \delta_{a}^{b}\, \delta_{i}^{j}\,\, \delta^{3} (x, y)\, . \ee
We begin by considering the simplest gauge covariant functions  on this phase space $\psym$ and impose them as constraints, as in (\ref{constraints}):
\be \label{con2} \mathcal{G}_{i} := \D_{a} \E^{a}_{i}  \simeq0; \quad  \V_{a}  := \E^{b}_{i} F_{ab}{}^{i} \simeq 0;\quad {\rm and} \quad  \S:= \textstyle{\f{1}{2}} \,\,\ez^{ij}{}_{k}\E^{a}_{i} \E^{b}_{j} F_{ab}{}^{k} \simeq 0 \, .\ee
 (The factor of $\f{1}{2}$ in the expression of $\S$  is for later convenience; see Section \ref{s4.2}.) As we noted in Section \ref{s1}, the first set, $\G =0$, is the Gauss constraint and  the canonical transformations it generates are the standard internal $SU(2)$ gauge rotations of the YM theory. The second and the third set, $\V_{a} \simeq 0,\,\, {\rm and} \,\,\S \simeq 0$  are called the \emph{vector} and \emph{scalar} constraint; they are new from the gauge theory perspective. 
As noted in Section \ref{s1},  one can verify that this is a system of  7 first class constraints \`a la Dirac. Since we are interested in a background independent theory, we are led to introduce  a linear combination of the three constraints as the Hamiltonian of the theory: 
\be \label{ham1} H_{\Lambda,\Sv,\N} (A,E) = \int_{\Sigma} \rmd^{3}x\,  \big( \Lambda^{i}\G_{i} + S^{a} \V_{a} + \N \S \big) \, \ee
using Lagrange multipliers $\Lambda^{i},  S^{a}$ and $\N$. The vector field $S^{a}$ is called the \emph{shift}, and  $\N$, a scalar density of weight $-1$ is called the \emph{lapse} in this framework. Note that because the integrand has density weight $1$, the integral is well defined without reference to a background volume element.  Note also that the Hamiltonian is  a low order polynomial in the basic canonical pair $(A_{a}^{i},\, \E^{a}_{i})$.  Therefore, the equations of motion it generates are gauge covariant and simple (see Section \ref{s4.1}). This is our background independent gauge theory with a set of first class constraints and precisely two true degrees of freedom. Interestingly this theory is completely equivalent to geometrodynamics discussed in Section \ref{s1} with an appropriate dictionary. 

The dictionary is the following. Let us first consider the recovery of Riemannian GR with signature +,+,+,+ from the gauge theory framework. Each electric field $\E^{a}_{i}$ provides a map from fields $\lambda^{i}$ taking values in  $su(2)$   to vector fields (with density weight $1$) $\tilde\lambda^{a}$ on $\Sigma$:\, $\lambda^{i} \to \tilde\lambda^{a} := \E^{a}_{i} \lambda^{i}$. Let us restrict ourselves to the generic case when the map is 1-1. Then $\E^{a}_{i}$ defines a +,+,+ metric $q_{ab}$ on $\Sigma$ via:\, $\tilde{\tilde{q}}\, q^{ab} = \qz^{ij} \E^{a}_{i} \E^{b}_{j}$ where $\tilde{\tilde{q}}$ is the determinant of $q_{ab}$.% 
\footnote{Recall that we have fixed orientation on $\Sigma$. Therefore $\Sigma$ admits a unique, metric independent Levi-Civita tensor density of weight $1$, which we will denote by  $\tilde\eta^{abc}$. Let $\epsilon_{abc}$ be the unique volume-form with the same orientation determined by the metric $q_{ab}$, i.e., satisfying $q^{am} q^{bn} q^{cp} \epsilon_{abc} \epsilon_{mnp} =3!$. Then $q^{\f{1}{2}} = \f{1}{3!}\,\t\eta^{abc} \epsilon_{abc}$. Alternatively, $\tilde{\tilde{q}} := {\textstyle{\f{1}{3!}}}\, \undertilde\eta_{abc} \ez^{ijk}\,\E^{a}_{i} \E^{b}_{j} \E^{c}_{k}$ and $q^{\f{1}{2}}$ is its positive square-root.}
Thus, we have recovered the configuration variable $q_{ab}$ of geometrodynamics. Through this dictionary, the electric field $\E^{a}_{i}$ acquires a dual interpretation as an orthonormal triad (with density weight $1$) for the metric $q_{ab}$. Next, to understand the interpretation of the gauge theory connection $A_{a}^{i}$ in terms of geometrodynamical fields, let us begin by noting that $\Sigma$ admits a unique derivative operator $\mathbf{D}$ that acts on both tensor and internal indices and annihilates $\E^{a}_{i}$ , i.e., satisfies $\mathbf{D}_{a} \E^{b}_{i} =0$. Let us focus on its action on fields with only internal indices. Then, $\Sigma$ admits  a connection 1-form $\Gamma_{a}^{i}$ and a 1-form $\pi_{a}^{i}$ which take values in $su(2)$ such that   
%\pi_{a}
\be \label{k1} \mathbf{D}_{a} \lambda_{i} = \partial_{a}\lambda_{i} + \ez_{ij}{}^{k}\, \Gamma_{a}^{j} \lambda_{k}\qquad {\rm and} \qquad (\mathbf{D}_{a} - \D_{a} )\, \lambda_{i} =: \ez_{ij}{}^{k} \,  \pi_{a}^{j}\,  \lambda_{k} \ee
for all $\lambda_{i}$. Therefore, $A_{a}^{i} = \Gamma_{a}^{i} - \pi_{a}^{i}$.  The field $\pi_{a}^{i}$ is essentially the extrinsic curvature $k_{ab}$ of geometrodynamics; more precisely,  
\be   k_{ab} := q^{-\f{1}{2}}\,\pi_{(a}^{i} q_{b)c}\,\E^{c}_{i}\, .\ee
 Recall that $k_{ab}$ determines the momentum $\P^{ab}$ conjugate to $q_{ab}$ via $\P^{ab} = - \epsilon\,q^{\f{1}{2}} \,(k^{ab} - k q^{ab})$. Thus, in the Riemannian sector a pair $(A_{a}^{i}, \,\E^{a}_{i})$ in $\psym$, (where $\E^{a}_{i}$ is non-degenerate)  determines the pair $(q_{ab}, \, \P^{ab})$ via
\be \label{map1}  \tilde{\tilde{q}} q^{ab} = \E^{a}_{i} \E^{b}_{j}\, \qz^{ij} \qquad {\rm and} \qquad\P^{ab} =  -\,\pi_{c}^{i}\, (q^{c(a} \E^{b)}_{i} -  \E^{c}_{i} q^{ab}) \equiv -\, q^{\f{1}{2}}\, (k^{ab} - k q^{ab}) \, . \ee
This completes the dictionary. We will see in Section \ref{s4.2} that $(q_{ab}, \,\P^{ab})$ satisfy the geometrodynamical constraints and evolution equations in the Riemannian signature if  $(A_{a}^{i}, \E^{a}_{i})$ satisfy the constraints (\ref{con2}) and the evolution equations generated by the Hamiltonian(\ref{ham1}). Thus, the map (\ref{map1}) provides us with a projection from the pairs $(A_{a}^{i},\, \E^{a}_{i})$ on $\psym$ to the pairs $(q_{ab},\, \P^{ab})$ of the geometrodynamical phase space $\psgeo$ that \\ (i) preserves the Poisson brackets (up to an overall constant); \\
 (ii) maps the constraint surface on $\psym$ to the constraint surface of $\psgeo$; and, \\ 
 (iii) sends dynamical trajectories on the constraint surface of $\psym$ to dynamical trajectories on the constraint surface of $\psgeo$.  \\
If $g_{ab}$ is a 4-d solution of Einstein's equation with initial data $(q_{ab},\, \P^{ab})$ on $\Sigma$, then $A_{a}^{i}$ turns out to have a natural geometrical interpretation: It is the pull-back of the self-dual part of the space-time connection that parallel transports unprimed spinors.  (${A}_{a}^{i} = \Gamma_{a}^{i} -  \pi_{a}^{i}$ is the pull-back to $\Sigma$ of the self-dual part of the full space-time connection on internal indices.)  {Therefore, as we will see in Section \ref{s4.3}, the gauge theory formulation is particularly well suited to describe Einstein dynamics  in the anti self-dual and self-dual sectors. Note, however, that the above Hamiltonian description on $\psym$ captures \emph{full Riemannian GR}; not just that of its (anti) self-dual sector.}

This may seem surprising at first  since, whereas all equations on $\psym$ are low order polynomials in basic variables, those on $\psgeo$ have a  complicated non-polynomial dependence. But this is 
simply because $(q_{ab}, \P^{ab})$ are complicated, non-polynomial functions of  $(A_{a}^{i}, \E^{a}_{i})$. Given that electroweak and strong interactions are described by gauge theories, it is interesting that equations of GR simplify considerably when the theory is also recast as a gauge theory --but now background independent, by necessity-- regarding Riemannian geometry as `emergent'.  Finally, note that (\ref{con2}) and  the evolution equations generated by the Hamiltonian of Eq. (\ref{ham1})  provide a slight generalization of Einstein's equations because they continue to be valid even when $\E^{a}_{i}$ fails to be 1-1, i.e., $q_{ab}$ becomes degenerate. We will return to this point in Section \ref{s5}.

Let us now consider Lorentzian general relativity.  In the Riemannian signature, equations simplified because the configuration variable $A_{a}^{i}$ turned out have the interpretation of  (the pull-back to $\Sigma$ of) the self-dual part of the spacetime connection in the final solution. The same is true in the Lorentzian signature. However, now the self-dual part of the connection is complex. Therefore we have to consider a complexification  $\psymc$ of $\psym$. The basic Poisson-bracket (\ref{fpb1}) is the same 
\be \label{fpb2} \{A_{a}^{i} (x),\, \E^{b}_{j}(y) \} \,  = \,  \delta_{a}^{b}\, \delta_{i}^{j}\,\, \delta^{3} (x, y) \,\ee
given by the natural extension of the symplectic structure on $\psym$ to $\psymc$, and the constraints continue to be given by (\ref{con2}). Since the Hamiltonian is a linear combination of these constraints, the equations of motion are also unchanged from those in the Riemannian case, but now refer to complex-valued  fields $A_{a}^{i}, \E^{a}_{i}$.  (The explicit form of the dynamical equations is given in Section \ref{s4.1}). It is clear from our discussion above that the Riemannian theory is recovered simply by restricting oneself to the real section $\psymr$ of $\psymc$ on which $A_{a}^{i}, \E^{a}_{i}$ are both real-valued. 

The real, Lorentzian  phase space  $\psyml$  of general relativity, on the other hand, corresponds to another `real' section of  $\psymc$ that is more subtle to specify. First, it is $\Eub^{a}_{i } := i \E^{a}_{i}$ --rather than the momentum $\E^{a}_{i}$--  that has the interpretation of the density-weighted orthonormal triad: {The positive definite 3-metric on $\Sigma$ is now given by  $q q^{ab} = \Eub^{a}_{i}\, \Eub^{b}_{j}\, q^{ij}$.}  To spell out the restriction on $A_{a}^{i}$,  let us, \emph{as before}, denote the derivative operator that annihilates $\Eub^{a}_{i}$  (and hence also $\E^{a}_{i}$)  by $\mathbf{D}$, and define the connection 1-form $\Gamma_{a}^{i}$ and the $su(2)$-valued  1-form $\pi_{a}^{i}$ via:
\be \label{k2} \mathbf{D}_{a} \lambda_{i} = \partial_{a}\lambda_{i} + \ez_{ij}{}^{k}\, \Gamma_{a}^{j}\, \lambda_{k}\qquad {\rm and} \qquad (\mathbf{D}_{a} - \D_{a} )\, \lambda_{i} =: \ez_{ij}{}^{k} \,  \pi_{a}^{j}\,  \lambda_{k} \ee
On the Lorentzian section, it is $\kub_{a}^{i} = -i \pi_{a}^{i}$ that is real-valued so that  $k_{ab}:= q^{-\f{1}{2}}\,\,\kub_{(a}^{i} q_{b) c}\, \Eub^{c}_{i}$ is real.  As in the Riemannian case, on dynamical trajectories $k_{ab}$ has the interpretation of the extrinsic curvature. To summarize, the real  Lorentzian section $\psyml$ of $\psymc$ is the one on which the pair  of fields $\Eub^{a}_{i}, \,\kub^{i}_{a}$ is real modulo gauge rotations --i.e., their imaginary parts, if any, are pure gauge in the sense that they can be can be removed by  the same internal rotation. 

The  dynamical flow on $\psymc$ (discussed in Section \ref{s4.1}) is tangential to this section (just as it is to the Riemannian section $\psymr$ on which $\E^{a}_{i}$ and $\pi_{a}^{i}$ are real). The dictionary is:
\ba \label{map2}  \tilde{\tilde{q}}\, q^{ab} &=& \Eub^{a}_{i}\, \Eub^{b}_{j}\, \qz^{ij}  \equiv - \E^{a}_{i} \E^{b}_{j}  \qz^{ij}  \qquad {\rm and} \nonumber\\
 \P^{ab} &=&  \kub_{c}^{i} (q^{c(a} \Eub^{b)}_{i} -  \Eub^{c}_{i} q^{ab})  \equiv  \,\pi_{c}^{i} (q^{c(a} \E^{b)}_{i} -  \E^{c}_{i} q^{ab}) \equiv q^{\f{1}{2}}\,(k^{ab} - kq^{ab}) \, .  \ea
We will see in Section \ref{s4.2} that this map provides us with a projection from the pairs $(A_{a}^{i},\, \E^{a}_{i})$ in $\psyml$  to the pairs $(q_{ab},\, \P^{ab})$ of the Lorentzian geometrodynamical phase space $\psgeo^{\rm Lor}$ that has the same three properties as in the Riemannian case: It \\    
(i) preserves the Poisson brackets (up to an overall constnat); \\
(ii) maps the constraint surface on $\psym$ to the constraint surface of $\psgeo$; and, \\ 
(iii) sends dynamical trajectories on the constraint surface of $\psym$ to dynamical trajectories on the constraint surface of $\psgeo$.\\ 
While in the Riemannian sector the self-dual connection $A_{a}^{i}$ is given by $A_{a}^{i} = \Gamma_{a}^{i} - \pi_{a}^{i}$ (with $\pi_{a}^{i}$ real), in the geometrodynamical variables, Lorentzian sector it is given $A_{a}^{i} = \Gamma_{a}^{i} - i \kub_{a}^{i}$ (with $\kub_{a}^{i}$ real).  

From the geometrodynamical perspective, as noted in (section 2 of chapter 8 of) \cite{aabook} the real Lorentzian section of the complexified phase space is the one on which the metric and the extrinsic curvature are real. This sector is  recovered from the projection map by restricting ourselves to the subspace of $\psymc$ on  which
\be \label{reality}  \Eub^{a}_{i}\Eub^{b}_{j}\, \qz^{ij}\,\,\, \hbox{\rm is positive definite} \qquad {\rm and} \qquad \dot{\Eub}^{(a}_{i} \Eub^{b)}_{j}\, \qz^{ij}\,\,\, \hbox{\rm is real}, \ee
using equations of motion generated by the Hamiltonian $H_{\Lambda,\Sv,\N}(A,E)$ discussed in Section \ref{s4.1} (with $\N$ real, and, say, $\Lambda=0, S^{a}=0$). These conditions ensure that the initial data  in geometrodynamics are real.  Evolution equations on the YM phase space preserve these reality conditions.  Note that these `Lorentzian reality conditions' are also low order-polynomials in $(A_{a}^{i},\, E^{a}_{i})$.  

To summarize, one can recover GR from a natural background independent gauge theory. (Results presented in this section have been extended to include the fields  --scalar, Dirac and YM-- that feature in the standard model \cite{art,aabook}.)  This procedure has the advantage that it leads to an enormous simplification of the constraint as well as evolution equations. In this formulation, there is no mention of a metric, its derivative operator that acts on tensor fields, or the extrinsic curvature, all of which are ubiquitous in the equations of geometrodynamics.  The Riemannian/pseudo-Riemannian geometry that underlies GR can be thought of as a secondary, emergent structure, albeit one that plays a central role both in classical GR, as well as in quantum field theory on a background space-time.\\ 

\emph{Remarks:}

1.   In our dictionary,  the Poisson bracket between the geometrodynamical variables is given by $\{q_{ab}(x),\, \P^{cd}(y) \} = 2\epsilon\, \delta_{(a}^{c}\, \delta_{b)}^{d}\, \delta^{3}(x,y)$.  The factor of $2$ comes from the fact that the triads $\E^{a}_{i}$ are `square roots' of the metric, and the $\epsilon$ comes from the fact that they directly determine the contravariant rather than the covariant metric, together with the $\epsilon$ in the relation $\P^{ab} = -\epsilon q^{\f{1}{2}}\, (k^{ab} - k q^{ab})$ between the momentum and the extrinsic curvature.

2.  In much of the LQG literature,  the Poisson bracket in the Lorentzian sector has an $i$ on the right side. Then the Riemannian and Lorentzian phase spaces, $\psymr$ and $\psyml$ do not arise as (appropriately defined) `real' sections of a \emph{single} gauge theory phase space $\psymc$.  In the present formulation they do. Technically, this point is trivial but conceptually it is a genuine advantage to have both sectors embedded in the same phase space. As we will see in Section \ref{s4.2},  thanks to this feature, GR dynamics in \emph{both} signatures gets encoded in a single Hamiltonian on the gauge theory phase space.

\section{Lie derivatives on fields with internal indices}
\label{s3}

While the constraint  and time-evolution equations are much simpler in the gauge theory framework than in geometrodynamics,  the disparity between `space-evolution' and `time-evolution' noted in Section \ref{s1} still persists \cite{alrev,ttbook,30years}. To bridge this gap, in this section we will make a detour to generalize the standard notion of  Lie derivatives in two steps.  In section \ref{s3.1} we recall  the notion of gauge covariant Lie derivative, and in section \ref{s3.2} we generalize it so that it is now defined with respect to vector fields that themselves carry an internal index.  In section \ref{s4} we will show that the first notion captures the canonical transformation generated by the vector constraint  --i.e., `space-evolution'-- and the second captures the canonical transformation generated by the scalar constraint  --i.e., `time-evolution'.

\subsection{Gauge covariant Lie derivatives}
\label{s3.1}

To bring out the new elements involved in the generalization, let us begin by collecting a few basic properties of Lie derivatives. While Lie derivatives descend from the action of diffeomorphisms generated by vector fields $V^{a}$ on $\Sigma$, their action on any tensor field is completely determined by the following 4 properties: (i) Linearity; (ii) Satisfaction of the Leibniz rule; (iii) Action on the ring of functions ($\l_{\vec V} f = V \vins \rmd f$); and, (iv) Commutativity with respect to the exterior derivative on functions ($\l_{\vec V}\, \rmd f = \rmd\, \l_{\vec V} f$).  In particular, these properties imply:
\be   \label{lie1}  \l_{\vec V}\, T_{a}{}^{b} = V^{c} \mfd_{c} T_{a}{}^{b} + T_{c}{}^{b} \mfd_{a} V^{c} - T_{a}{}^{c} \mfd_{c} V^{b}\, . \ee
where $\mfd$ is \emph{any} derivative operator on $\Sigma$ that is torsion-free (i.e. satisfies $\mfd_{[a} \mfd_{b]} f =0$ for all functions $f$), which could be taken to be a flat derivative operator $\partial$.  Let us make this choice for simplicity, so that we have:
\label{lie1prime} \be \l_{\vec V}\, T_{a}{}^{b} = V^{c} \partial_{c} T_{a}{}^{b} + T_{c}{}^{b} \partial_{a} V^{c} - T_{a}{}^{c} \partial_{c} V^{b}\, . \ee
Secondly, the commutator of the action of Lie derivatives w.r.t. two vector fields $U^{a}$   and $V^{a}$ is given by their Lie derivative:
\be \label{com1}  [\l_{\vec U},\,  \l_{\vec V}]\, T_{a}{}^{b} = \l_{\vec W} T_{a}{}^{b} \qquad {\rm where} \qquad W^{a} = \l_{\vec U} V^{a}\, . \ee

One can extend  the notion of Lie derivatives to \emph{generalized tensor fields} --i.e. tensor fields with both tensor and internal indices that transform covariantly under gauge transformations--  in two different ways. The most obvious strategy is to regard the internal indices \emph{as scalars,}  so that $\l_{\vec V}$ simply ignores them.  We can again use any torsion-free derivative operator $\mfd$ on tensor indices. For simplicity let us again use a flat $\partial$. Then one has 
\be \label{lie2}
 \l_{\vec V} \,T_{ai}{}^{bj} := V^{c} \partial_{c} T_{ai}{}^{bj} + T_{ci}{}^{bj} \partial_{a} V^{c} - 
T_{ai}{}^{cj} \partial_{c} V^{b}\, , \ee
where the action of $\partial$ also ignores the internal indices (so that, $\partial$  acts on $T_{i}{}^{j}$, for example, as though it were a function, whence $\l_{\vec V}T_{i}{}^{j} = V^{a}\partial_{a} T_{i}{}^{j} $). In particular,  the action of $ \l_{\vec V}$ on the basic canonical variables is given by:
\ba \label{lA}  \l_{\vec V} A_{a}{}^{i} &=& V^{b}\, \partial_{b} A_{a}{}^{i} + A_{b}{}^{i} \,\partial_{a} V^{b} \\ 
 \label{lE} \l_{\vec V}\, \E^{a}_{i} &=&  V^{b} \,\partial_{b} \E^{a}_{i} - \E^{b}_{i}\, \partial_{b} V^{a} + \E^{a}_{i}\,\partial_{c} V^{c}\, ,  \label{lE}\ea
where the last term in (\ref{lE}) arises because $\E^{a}_{i}$ has density weight $1$. This is the most commonly used strategy in the LQG literature. It has several attractive features: (a) properties (i) - (iv) are carried over;  (b) calculations are easier, being closer to those involving the Lie derivatives on just tensor indices;  (c) one can again use any torsion-free derivative operator $\mfd$ in place of $\partial$; and, (d)  the property (\ref{com1}) extends to fields $T_{a_{1}  \ldots a_{m}  i_{1} \ldots i_{n} }{}^{b_{1} \ldots b_{k} j_{1} \ldots j_{\ell}}$ with both tensor and internal indices. However even when  $T_{ai}{}^{bj}$ is a  generalized tensor field  --and hence, \emph{gauge covariant} in internal indices--  the result of the action,  $\l_{\vec V} T_{ai}{}^{bj}$,  is \emph{no longer gauge covariant.} 

The second natural possibility is to use a \emph{gauge covariant} (GC) Lie derivative $\gcl$ on generalized tensors fields. This requires us to make a choice of a derivative operator $\D$ on \emph{internal} indices which, however, can be extended to tensor indices using \emph{any} torsion-free derivative operator $\mfd$. Let us again choose it to be flat for concreteness. Then the  definition of $\gcl_{\vec V}$ has the same form as  (\ref{lie1}),  with $\mfd$ replaced by $\D$  while acting on internal indices. Thus, one sets:
\ba \label{lie3}
\gcl_{\vec V} T_{ai}{}^{bj} &:=& V^{c} \D_{c} T_{ai}{}^{bj} + T_{ci}{}^{bj} \partial_{a} V^{c} - 
T_{ai}{}^{cj} \partial_{c} V^{b} \nonumber \\
&=& V^{c} \partial_{c} T_{ai}{}^{bj} + \ez_{ik}{}^{\ell} V^{c} A_{c}^{k} T_{a\ell}{}^{bj} - \ez_{\ell k}{}^{j} V^{c} A_{c}^{k} T_{ai}{}^{b\ell} + T_{ci}{}^{bj} \partial_{a} V^{c} - T_{ai}{}^{cj} \partial_{c} V^{b} 
. \ea
This operator $\gcl_{\vec V}$  has the following features: \\
(1) Properties (i) - (iv) are carried over. In addition, we have $\gcl_{\vec V} T_{i} = V\vins \D T_{i}$.\\
(2) If $T_{ai}{}^{bj}$ is a generalized tensor field, so is $ \gcl_{\vec V}\,T_{ai}{}^{bj}$;  the operation now preserves gauge covariance.\\
(3) While on tensor indices we can use any torsion-free $\mfd$ (in place of $\partial$), on internal indices we have to use a fixed $\D$.\\
 (4) While property (\ref{com1}) continues to hold for action of $\gcl$ on tensor fields, it is modified in a crucial way for fields with internal indices. In particular, we have:
\be \label{com2}  [\gcl_{\vec U},\,  \gcl_{\vec V}]\, T_{i} = \gcl_{\vec W} T_{i}  + U^{a}V^{b} \ez_{ij}{}^{k} \,F_{ab}{}^{j}\, T_{k}  \qquad  \hbox{\rm where, again} \qquad W^{a} := \gcl_{\vec U} V^{a}  \equiv \l_{\vec U} V^{a}\ee
and $F_{ab}{}^{j}$ is the curvature of the chosen $\D$.  Thus the commutator of the actions of two vector fields is now a linear combination of the gauge covariant Lie derivative \emph{and} an infinitesimal internal gauge rotation. (Furthermore,  the generator of the gauge rotation on $\psym$ is now a `q-number' field in that it depends on (the phase space variable) $F_{ab}{}^{j}$.) Nonetheless, $\gcl_{\vec V}$ has a natural interpretation in terms of lifts of  the vector field $V^{a}$ on $\Sigma$ to the (appropriate, associated) $SU(2)$ bundle over $\Sigma$, defined by the horizontal cross-sections selected by the connection 1-form $A_{a}^{i}$ \cite{jackiw}.

Let us conclude by writing down the action of the GC Lie derivative on phase space variables.  Since curvature $F_{ab}{}^{i}$ and the triad $E^{a}_{i}$ are generalized tensors -- and therefore transform covariantly under local gauge rotations-- the action is directly given by (\ref{lie3}):
\ba \label{gclF}  \gcl_{\vec V} F_{ab}{}^{i} &=& V^{c}\, \D_{c} F_{ab}{}^{i} + F_{cb}{}^{i} \,\partial_{a} V^{c} +F_{ac}{}^{i}\, \partial_{b} V^{c} \\ 
 \gcl_{\vec V}\, \E^{a}_{i} &=&  V^{c} \,\D_{c} \E^{a}_{i} - \E^{c}_{i}\, \partial_{c} V^{a} + \E^{a}_{i}\,\partial_{c} V^{c}\, ,  \label{gclE}\ea
where, again, the last term in (\ref{gclE}) arises because $\E^{a}_{i}$ has density weight $1$. (The  final result is again independent of the torsion-free derivative operator on tensors used in extending the action of $\D$ to tensor indices.)  We cannot directly use (\ref{lie3}) to find the action of $\gcl_{\vec V}$ on the connection 1-form $A_{a}^{i}$ because it does not transform covariantly under gauge transformations.  However, since $\D_{a} T_{i}$ does transform covariantly,  to find the action of $\gcl_{\vec V}$ on $A_{a}^{i}$ we can use (\ref{lie3}) to calculate the action on $T_{i}$ of the commutator between $\gcl_{\vec V}$ and $\D$. One finds:
\be \big(\gcl_{\vec V}\, \D_{a} - \D_{a}\, \gcl_{\vec V} \big) T_{i} \,=\, \ez_{ij}{}^{k}\, V^{b}\, F_{ba}{}^{j} T_{k}  \ee
Therefore, in view of the definition (\ref{A}) of $A_{a}^{i}$  we are led to define the \emph{gauge covariant Lie derivative of the connection 1-form} $A_{a}^{i}$ as:
\be \label{gclA} \gcl_{\vec V} A_{a}^{i} := V^{b}F_{ba}{}^{i} \, .\ee
One can check the consistency of this definition by first expressing  $F_{ab}{}^{i}$ in terms of $A_{a}^{i}$ using (\ref{A}), and  calculating $\gcl_{\vec V} F_{ab}{}^{i}$ using (\ref{gclA}).  One recovers precisely the expression (\ref{gclF}) above. The expression (\ref{gclA}) of the gauge covariant Lie derivative of the connection 1-form can also be obtained by first lifting  $V^{a}$ horizontally from $\Sigma$ to the principal fibre bundle (where the horizontal lift is defined by $A_{a}^{i}$) and using the diffeomorphism generated by the lift on the bundle. 

Eqs. (\ref{gclA}) and (\ref{gclE}) will be directly useful in spelling out the geometrical meaning of the canonical transformation generated by the vector constraint $\V_{a} =0$ in Section \ref{s4}. We have spelled out the basic structure of GC Lie derivatives because these considerations will directly motivative the generalization introduced in the next sub-section. \\

\emph{Remarks:} \\
\indent 1. Following the standard practice in gauge theories, we have used the connection 1-form $A_{a}^{i}$ as the configuration variable. However,  in place of $A_{a}^{i}$, conceptually it is more appropriate to use the derivative operators $\D$ that act on fields with internal indices. Then the entire framework becomes gauge covariant. The configuration space $\mathcal{C}$ is an (infinite dimensional) affine space of the derivative operators $\D$; tangent vectors to $\mathcal{C}$ are represented by infinitesimal differences $\delta\D$ between two derivative operators --and hence by $\delta A_{a}^{i}$ in the present terminology-- and cotangent vectors continue to be the electric fields $\E^{a}_{i}$ which act on tangent vectors via $\delta A_{a}^{i} \to \int_{\Sigma} \rmd^{3} x\,\E^{a}_{i}\, \delta A_{a}^{i}$.  Both $\delta A_{a}^{i}$ and $\E^{a}_{i}$ transform covariantly under local gauge transformations. The phase space is the cotangent bundle over $\mathcal{C}$. This shift in the choice of configuration space does not change any of the results. But it serves to remove the awkwardness that can arise --e.g. in the definition of $\gcl_{\rm V} A_{a}^{i}$--  because $A_{a}^{i}$ is not gauge covariant.

2. Similarly, following the common practice, we denoted conjugate momenta  by electric fields,  $\E^{a}_{i}$, that are vector fields with density weight 1. The density weight brings in the extra term $\E^{a}_{i} \partial_{c}V^{c}$  in the expressions (\ref{lE})  and  (\ref{gclE})  of the Lie derivative and the GC Lie derivative of $\E^{a}_{i}$.  However, it is more natural to represent the momentum by an $su(2)$-valued 2-form $e_{ab}{}^{i} := \undertilde{\eta}_{abc}\, \E^{c}_{j} \,\qz^{ij}$. Then the symplectic structure becomes: 
\be \Omega\mid_{(\D,\, e)} (\delta_{1}, \delta_{2}) = \int_{\Sigma} \big(\delta_{1} A^{i} \wedge\delta_{2}  e^{j}\, -\, \delta_{1} e^{i} \wedge\delta_{2}  A^{j}\big) \qz_{ij} , \ee
where $\delta \equiv (\delta{A}_{a}^{i}, \delta e^{i}_{ab})$ denotes a tangent vector to $\psym$ at the point $(\D, e_{ab}^{i})$. And  (\ref{gclE}) is replaced by the familiar expression of the action of the gauge covariant Lie derivative of an $su(2)$-valued 2-form:
\be \gcl_{\vec V}\, e_{ab}{}^{i} = V^{c}\D_{c} e_{ab}{}^{i} + e_{cb}^{i} \partial_{a} V^{c} + e_{ac}^{i} \partial_{b} V^{c} . \ee
The `extra term' $\t{E}^{a}_{i} \partial_{c}V^{c}$ emerges when we pass from $e_{ab}^{i}$ to $\E^{a}_{i}$

\subsection{Generalized gauge covariant Lie derivatives}
\label{s3.2}

So far we have restricted ourselves to Lie derivatives of tensor fields and generalized tensor fields with respect to vector fields $V^{a}$. But now that we have fields carrying internal indices, it is natural to extend the notion of a Lie derivative with respect to vector fields $V^{a}_{i}$\, \emph{that themselves carry internal indices}.  The discussion of Section \ref{s3.1} suggests a natural avenue to define these generalized gauge covariant (GGC) Lie derivatives $\ggcl_{\vec V_{k}}$: We can just replace $\partial_{a} V^{b}$ in  (\ref{lie3}) with $\D_{a} V^{b}_{k}$ thereby ensuring gauge covariance of the result. Thus we will set
\be \label{lie4}
\ggcl_{\vec V_{k}} \,T_{ai}{}^{bj} := V^{c}_{k}\, \D_{c} T_{ai}{}^{bj} + T_{ci}{}^{bj}\, \D_{a} V^{c}_{k} - 
T_{ai}{}^{cj} \,\D_{c} V^{b}_{k} \ee
It is obvious that for ordinary vector fields $V^{a}$, the definition reduces to the standard GC Lie derivatives. The key difference is that fields $T_{ai}{}^{bj}$ are now mapped to generalized tensor fields  that carry \emph{an additional internal index.} (In some ways this action is analogous to that of a derivative operators $\mfd$ acting on tensor fields on $\Sigma$: $T_{a}{}^{b}$  is mapped to the tensor field $\mfd_{c} T_{a}{}^{b}$ that carries an additional index $c$.)  However, the operator has the same properties as those satisfied by $\gcl_{\vec V}$, albeit natural modifications:\\
 ($1^\prime$) The obvious generalizations of properties (i) - (iv) hold. That is, the action of $\ggcl_{\vec V_{k}}$ on generalized tensor fields is linear; satisfies  the Leibniz rule; maps functions $f$ on $\Sigma$ to a field with an internal index $k$ via  $\ggcl_{\vec V_{k}} f = V_{k} \vins \rmd f$; and, commutes with the the operation of exterior derivative on functions; \,\,\,\,$\ggcl_{\vec V_{k}} \, \rmd f  \equiv \ggcl_{\vec V_{k}} \, \D\, f =  \D\, \ggcl_{\vec V_{k}} f$. 
 Finally, as with $\gcl_{\vec V}$, we have the following additional property: $\ggcl_{\vec V_{k}} T_{i} = V_{k} \vins \D T_{i}$. \\
  ($2^{\prime}$) If $T_{ai}{}^{bj}$ is a generalized tensor field, so is $ \ggcl_{\vec V_{k}}\,T_{ai}{}^{bj}$;  the operation again preserves gauge covariance.\\
($3^{\prime}$) While on tensor indices we can use any torsion-free $\mfd$, on internal indices we have to use a fixed $\D$.\\
($4^{\prime}$) Direct analog of property (\ref{com2}) holds after making obvious changes to accommodate the fact that $V^{a}_{k}$ now carries an internal index:
\be \label{com3}  [\ggcl_{\vec U_{i}},\,  \ggcl_{\vec V_{j}}]\, T_{k} = \ggcl_{\vec W_{ij}} T_{k}  + \ez_{kl}{}^{m}\, U^{a}_{i} \,V^{b}_{j} \,F_{ab}{}^{l}\, T_{m} \ee 
where $F_{ab}{}^{j}$ is, as before, the curvature of the chosen $\D$ and 
\be W^{a}_{ij} := \ggcl_{\vec U_{i}} V^{a}_{j}  =  U^{b}_{i} \D_{b} V^{a}_{j}  -  V^{b}_{j} \D_{b} U^{a}_{i} \, . \ee
On the canonical pair $(A_{a}^{i}, \E^{a}_{i})$ the action of $\ggcl_{\vec V_{k}}$ is the obvious 
generalization of (\ref{gclA}) and (\ref{gclE}):
\ba \label{ggclA} \ggcl_{\vec V_{k}} A_{a}^{i} &=& V^{b}_{k}\,F_{ba}{}^{i} \\
\ggcl_{\vec V_{k}}\, \E^{a}_{i} &=&  V^{c}_{k} \,\D_{c} \E^{a}_{i} - \E^{c}_{i}\, \D_{c} V^{a}_{k} + \E^{a}_{i}\, \D_{c} V^{c}_{k}\, ,  \label{ggclE}\ea
One can verify that, together with the expression (\ref{A}) of  $F_{ab}{}^{i}$, Eq. (\ref{ggclA}) implies that the action of  $\ggcl_{\vec V_{k}}$ on  $F_{ab}{}^{i}$ is the expected one from (\ref{lie4}):
\be \label{fgclF}   \ggcl_{{\vec V}_{k} } F_{ab}{}^{i} = V^{c}{}_{k}\, \D_{c} F_{ab}{}^{i} + F_{cb}{}^{i} \,\D_{a} V^{c}_{k} +F_{ac}{}^{i}\, \D_{b} V^{c}_{k}\, .  \ee

\emph{Remark:} In the abstract index notation  \`a la Penrose \cite{rp, rpwr} and the `operational' view to tensor calculus \`a la Penrose and Geroch \cite{rg}, the action of the Lie derivative $\l_{\vec V}$ on tensor fields is completely determined by its properties (i)-(iv) noted in the beginning of Section \ref{s3.1}.  Therefore,  in the Penrose-Geroch framework,  one can simply \emph{define} $\l_{\vec V}$ by these properties, and \emph{then} show  that the action corresponds to the infinitesimal changes in the tensor fields under the action of the 1-parameter family of diffeomorphisms generated by $V^{a}$. Since (i) -(iv)  are the properties that are the ones that are used most commonly in all calculations, this approach provides an `operational' definition $\l_{\vec V}$. 

One can extend the abstract index notation  \`a la Penrose \cite{rp, rpwr} and the `operational' view to tensor calculus \`a la Penrose and Geroch \cite{rg} to generalized tensor fields that have both internal and tensor indices \cite{ahm}. (For a summary, see Appendix A.)  Then the  action of the GC Lie derivative $\gcl_{\vec V}$ is completely specified by properties listed under (1) in Section \ref{s3.1}, and of the GGC Lie derivative  $\ggcl_{\vec V}$ by properties listed under ($1^{\prime}$) in this sub-section.  Therefore, we can adopt these properties as the definition of the operators  $\gcl_{\vec V}$ and  $\ggcl_{\vec V}$. Now, in the case of $\gcl_{\vec V}$, one can then show that the action can be interpreted in terms of diffeomorphisms,  but now generated by the horizontal lifts of $V^{a}$  to the appropriate $SU(2)$ bundle over $\Sigma$. Can one show that our `operationally defined' $\ggcl_{\vec V_{k}}$ also admits a similar interpretation? This interesting mathematical issue remains open. Appendix A shows that the GGC Lie derivatives satisfy the Jacobi identity,  suggesting that an interpretation in terms of a group action should exist, but could involve bundles over $\Sigma$ with infinite dimensional fibers.  The existence of the Jacobi identity may also be of interest to the `double-copy' community.

\section{The gravitational dynamics on ${\mathbf \Gamma_{\bf {YM}}}$}
%{The gravitational evolution equations in connection-dynamics}
\label{s4}

In this section, we will combine the results of Sections \ref{s2} and \ref{s3} to bring `time-evolution' on the same footing as `space-evolution'. This step also makes the structure of constraints more transparent and simplifies calculations, e.g., of the constraint algebra.  The section is divided into 3 parts. In the first we discuss the canonical transformations generated by the vector and the scalar constraints on $\psymc$. These dynamical equations are rather simple on $\psymc$.  In the second part we show that the dictionary given in Section \ref{s2} enables one to recover the complicated equations of geometrodynamics  in both the Riemannian and the Lorentzian sectors. In the third part we discuss some of the important implications of the gauge theory formulation of Einstein's equations.

\subsection{Canonical transformations generated by constraints on $\mathbb{C}{\mathbf \Gamma_{\bf {YM}}}$}
\label{s4.1}

Let us begin with the phase space $\psymc$ consisting of complex-valued pairs $(A_{a}^{i},\, \E^{a}_{i})$, subject to the gauge theory Poisson brackets  (\ref{fpb2}):   
\be \label{fpb3} 
 \{A_{a}^{i} (x),\, \E^{b}_{j} (y) \} \, = \, \delta_{a}^{b}\, \delta_{i}^{j}\, \delta^{3}(x,y) \, . \ee
Let us begin with the vector constraint  $\V_{a}$.  When we smear it  with the shift vector field $S^{a}$ (and set the generator of internal rotations $\Lambda^{i}$ and the lapse $\N$ to zero) we obtain the Hamiltonian 
\be  \label{hamvector} H_{0,\Sv,0} (A, \E) = \int_{\Sigma}  \rmd^{3}x\, S^{a} \V_{a} \, {\rmd}^{3} x =  \int_{\Sigma} \rmd^{3}x\, S^{a} \E^{b}_{i} F_{ab}^{i}  \ee
on $\psym$. Using (\ref{fpb3})  it immediately follows that the infinitesimal canonical transformation generated by $H_{0,\Sv,0} (A,E)$ is given by
\ba \label{sevo} \dot{A}_{a}^{i} &=&   S^{b}F_{ba}^{i} \,\equiv\, \gcl_{\Sv} A_{a}^{i}, \,\,\, {\rm whence}\,\,\,
       \dot{F}_{ab}{}^{i} =  \gcl_{\Sv} F_{ab}{}^{i}, \quad {\rm and}, \nonumber  \\
\dot{\E}^{a}_{i}  &=& 2 \D_{b} (S^{[b} \E^{a]_{i}}) = \gcl_{\Sv} \E^{a}_{i} - (S^{a}\,\D_{b}\E^{b}_{i}) 
\ea
where we have used properties of the gauge covariant Lie derivative $\gcl_{\Sv}$ discussed in Section \ref{s3.1}. Thus, the Hamiltonian vector field $X_{\Sv}$ generated by $H_{0,\Sv,0} (A,E)$ is given by:
\be \label{hamS} X_{\Sv} =  \big(\gcl_{\Sv} A_{a}^{i}\big)\, \f{\delta}{\delta A_{a}^{i}} + \big( \gcl_{\Sv} \E^{a}_{i} - (S^{a\,}\D_{b}\E^{b}_{i})\big)\f{\delta}{\delta \E_{i}^{a}}  \ee
Its action on functions only of connections is the same as that of the infinitesimal gauge covariant diffeomorphism generated by $V^{a}$ everywhere on $\psym$. This property is useful  in quantum theory since one uses the connection representation (i.e. spin network states) in LQG.  Because $H_{0,\Sv,0}$ is gauge invariant, it follows that the vector field $X_{\Sv}$ is tangential to the Gauss-constraint surface defined by $\G_{i} \equiv \D_{a}\E^{a}_{i} =0$. The action of $X_{\Sv}$ on the restriction of any phase space function $f(A,E)$ to this surface is also the same as that of the infinitesimal GC diffeomorphism generated by $V^{a}$. Note that the result of the action of $X_{\Sv}$ is gauge covariant: it involves the \emph{gauge covariant Lie derivative} $\gcl_{\Sv}$, and not the ordinary Lie derivative $\l_{\Sv}$. On tensor fields, such as the ADM variables $(q_{ab},\, \P^{ab})$, constructed from $(A_{a}^{i}, \E^{b}_{j})$ the two operators coincide but on generalized tensor fields --such as the curvature $F_{ab}{}^{i}$-- constructed from the canonical pair, the infinitesimal change is again gauge covariant precisely because the action is via $\gcl_{\Sv}$ rather than $\l_{\Sv}$. 

The diffeomorphism constraint $\t{C}^{a}$ on the geometrodynamical phase space $\psgeo$ can be translated to the gauge theory $\psym$ using the dictionary between the two sets of variables. Modulo an overall constant, it is given by choosing the shift to be $S^{a}$, setting the generator of internal rotations to $\Lambda^{i} = -S^{a}\!A_{a}^{i}$, and vanishing lapse, $\N\! =0$:
\be H^{\rm diff}_{\Lambda, \Sv, 0} (A,\E) = \int_{\Sigma}  \rmd^{3}x\, \big(\Lambda^{i}\, \G_{i} + S^{a} \V_{a}\big)\, =\, \int_{\Sigma}   \rmd^{3}x\, \big(\! - \!S^{a}\!A_{a}^{i}\, \D_{b}\E^{b}_{i}  + S^{a} \E^{b}_{i} F_{ab}{}^{i}\big) \, .
\ee
Note that, because of the presence of the term $S^{a}A_{a}^{i} $,  the function $H^{\rm diff}_{\Lambda, \Sv, 0}$ is not gauge invariant. Therefore the canonical transformation it generates is not gauge covariant. We now have:
\be \dot{A}_{a}^{i} =  \l_{\Sv} A_{a}^{i}\qquad {\rm and}, \qquad \dot{\E}^{a}_{i}  = \l_{\Sv} \E^{a}_{i} \, ,\ee
with the standard Lie-derivative $\l_{\Sv}$ whose action fails to be gauge covariant (as we saw in Section \ref{s3.1}). In the LQG literature, since the gauge theory description was first arrived at starting from geometrodynamics, one generally considers this diffeomorphism constraint. From the gauge theory perspective, on the other hand, it is more natural to put gauge covariance at the forefront and consider the vector constraint.\\

Let us next consider the canonical transformation generated by the scalar constraint $H_{0,0,\N}$:
%\footnote{The factor of ${\textstyle{\f{1}{2}}}$ ensures that in the Lorentzian, asymptotically flat case, the numerical value of  $H_{0,0,\N}$ with the accompanying surface term is the ADM energy rather than a multiple thereof. This is because, on solutions to field equations, the integrand is equivalent to $G_{ab} n^{a}n^{a}$, where $n^{a}$ is the unit normal to the Cauchy slice.  This issue is further discussed below.}
%
\be \label{hamscalar} H_{0,0,\N} (A,\E) = \int_{\Sigma} \rmd^{3}x\,\, \N {\S}  := \int_{\Sigma} \rmd^{3}x\,\, {\textstyle{\frac{1}{2}}}\, \N \E^{a}_{i} \E^{b}_{j}\, F_{ab}{}^{k}\,  \ez^{ij}{}_{k} \ee
Comparison between the expressions (\ref{hamvector}) and (\ref{hamscalar})  of Hamiltonians defined by the vector and the scalar constraint suggests the introduction of an $su(2)$-valued vector field $N^{a}_{i}$ via
\be \label{Nai}  N^{a}_{i} := \N\, \E^{a}_{i}.  \ee
We can now use the expression (\ref{fpb3}) of Poisson brackets to calculate the infinitesimal canonical transformation generated by $H_{0,0,\N}$:
\ba \label{evo1} \dot{A}_{a}^{i} &=& - \N\E^{b}_{j}\ F_{ba}{}^{k}\, \ez^{\,ij}{}_{k}  \,\equiv\, -(\ggcl_{\Nv_{j}} A_{a}^{k}) \,\ez^{\,ij}{}_{k}, \,\,\, {\rm whence}\,\,\,
       \dot{F}_{ab}{}^{i} = - (\ggcl_{\Nv_{j}} F_{ab}{}^{k})\, \ez^{\,ij}{}_{k},  \quad {\rm and}, \nonumber \\
  \dot{\E}^{a}_{i}  &=&  \D_{b} (N_{j}^{[b} \E^{a]}_{k}\, \ez_{i}{}^{jk}) = {\textstyle{\f{1}{2}}} \,(\ggcl_{\Nv_{j}} \E^{a}_{k})\, \ez_{i}{}^{jk} - {\textstyle{\f{1}{2}}} \, N^{a}_{j} (\D_{b}\E^{b}_{k})\, 
 \ez_{i}{}^{jk} \, ,
\ea
where we have used the properties of the generalized gauge covariant  derivatives from Section \ref{s3.2}. Therefore, the corresponding Hamiltonian vector field is given by
\ba \label{hamN} X_{\N} &=&  -\ez^{ij}{}_{k}\big(\ggcl_{\Nv_{j}} A_{a}^{k}\big)\, \f{\delta}{\delta A_{a}^{i}} +{\textstyle{\f{1}{2}}}\,\ez_{i}{}^{jk} \Big(\ggcl_{\Nv_{j}} \E^{a}_{k} - (N^{a}{}_{j}\D_{b}\E^{b}_{k})\Big)\f{\delta}{\delta \E_{i}^{a}} \, , \nonumber \, \\
&\simeq& -\ez^{ij}{}_{k}\big(\ggcl_{\Nv_{j}} A_{a}^{k}\big)\, \f{\delta}{\delta A_{a}^{i}} +{\textstyle{\f{1}{2}}}\,\ez_{i}{}^{jk} \big(\ggcl_{\Nv_{j}} \E^{a}_{k} \big)\f{\delta}{\delta \E_{i}^{a}} 
 \ea
where the last equality holds on the Gauss constraint surface $\t\G \simeq 0$. Thus, as highlighted in the abstract, \emph{`time evolution'} is now expressed as a GGC Lie derivative \emph{along a space-like} $N^{a}_{i}$ on the Gauss constraint surface of the full phase space $\psymc$.  $N^{a}_{i} = \N \E^{a}_{i}$ is analogous to the shift vector field $S^{a}$ but  is a `q-number', Lie-algebra valued vector field because of its dependence on the dynamical variable $\E^{a}_{i}$. That is why we refer to $N^{a}_{i}$  as \emph{`an electric shift'}.  
Recall that in LQG quantum states are represented as functions of connections. On these functions, now the action of $X_{\N}$  has a direct geometrical meaning in terms of the GGC Lie derivative. (Recall that the action of $X_{\Sv}$ has a very similar geometrical meaning in terms of the GC Lie derivative.) This fact has motivated the recent definitions of the scalar constraint operator in LQG. (For a first attempt, see \cite{laddha}; a summary of the current status is given in Remark 3 of Section \ref{s5}.)

Next, note that the vector and the scalar constraint can be naturally combined if one defines vector fields $N^{a}{}^{j}{}_{i}$ and $S^{a}{}^{j}{}_{i}$, with internal indices, using the structure constant $\ez^{ij}{}_{k}$ and the Cartan-Killing metric $\qz_{ij}$:
\be \label{NSij} N^{a}{}^{j}{}_{k} :=  \ez^{ij}{}_{k} \, N^{a}_{i}\,=\, \ez^{ij}{}_{k}\,  \N \E^{a}_{i}\,  \quad {\rm and} \quad
 S^{a}{}^{j}{}_{k} := S^{a} \qz^{j}{}_{k} \equiv S^{a} \delta^{j}{}_{k} \, .\ee
The vector and the scalar constraints then blend together to yield
\be \label{blending2} H_{0,\Sv, \N}(A, \E) = \int_{\Sigma} \,\rmd^{3} x \big({\textstyle{\f{1}{2}}} N^{a}{}^{j}_{k} + S^{a}{}{}^{j}_{k}\big) \E^{b}_{j} \,F_{ab}{}^{k}  \ee
and the combined Hamiltonian generates the infinitesimal canonical transformations 
\ba \label{evo2} \dot{A}_{a}^{i}  \equiv \{A_{a}^{i}\, , H_{0,\Sv, \N}\} &=& \ggcl_{\Nv{}^{i}{}_{j} + \Sv{}^{i}{}_{j}}  A_{a}^{j}, \qquad {\rm and}\nonumber \\
\dot{E}^{a}_{i}  \equiv \{\E_{i}^{a}\, , H_{0,\Sv, \N}\}&=& \ggcl_{\f{1}{2}\Nv^{j}{}_{i} + \Sv^{j}{}_{i}}  E^{a}_{j}\, -\, \big( \textstyle{\f{1}{2}}\N^{a\,j}{}_{i} + \S^{a\,j}{}_{i}\big)  \D_{b}\E^{b}_{j}\, , \ea
where the `dot' now refers to the `time' defined by the lapse $\N$ \emph{together with} the shift 
$S^{a}$. (Note that we could replace $-\ggcl_{\Nv_{j}}\, \ez^{ij}_{k}$ with $\ggcl_{\Nv^{i}{}_{k}}$ because $\D_{a} \ez^{ij}{}_{k} =0$ (and $\D_{a} \qz_{ij} =0$).)  The mismatch of a factor of 2 between the time evolution of $A_{a}^{i}$ and $\E^{a}_{i}$ arises because $N^{a}{}^{j}{}_{i}$ is a `q-number' field:  the Poisson bracket of this `electric shift'  with $A_{a}^{i}$ is non-zero, while its Poisson bracket with $\E^{a}_{i}$ vanishes. Finally, the Hamiltonian vector field generated by the vector and the scalar constraints acquires the compact form:
\ba \label{hamSN} X_{\Sv,\N} &=&  \Big(\ggcl_{\Nv^{j}{}_{i} + \Sv^{j}{}_{i}} \,\, A_{a}^{i}\Big)\,  \f{\delta}{\delta A_{a}^{j}} + \Big( \ggcl_{\f{1}{2}\Nv^{j}{}_{i} + \Sv^{j}{}_{i}} \,\, E^{a}_{j}\, -\, ( {\textstyle{\f{1}{2}}}\Nv^{j}{}_{i} + \Sv^{j}{}_{i})  \D_{b}\E^{b}_{j}\Big)\, \f{\delta}{\delta \E_{i}^{a}} \nonumber \\
&\simeq& \Big(\ggcl_{\Nv^{j}{}_{i} + \Sv^{j}{}_{i}} \,\, A_{a}^{i}\Big)\,  \f{\delta}{\delta A_{a}^{j}} + \Big( \ggcl_{\f{1}{2}\Nv^{j}{}_{i} + \Sv^{j}_{i}} \,\, E^{a}_{j}\Big)\,\f{\delta}{\delta \E_{i}^{a}}    \, 
\ea
where the second equality holds on the Gauss constraint surface on which $\G_{i} \equiv \D_{a}E^{a}_{i}= 0$. Note that  $X_{\Sv,\N}$ is tangential to this surface because $H_{0,\Sv,\N}$ is gauge invariant. On this surface, \emph{time evolution} --i.e, the action of the combined Hamiltonian-- can be expressed \emph{entirely} in terms of GGC Lie derivatives along the \emph{spatial} electric shift. \\
\goodbreak
\emph{Remarks:}

1. At first it may seem that the evolution equations (\ref{evo2}) imply that the `time' derivative of $A_{a}^{i}$ depends only on $A_{a}^{i}$ and that of $\E^{a}_{i}$ depends only on $\E^{a}_{i}$, i.e., the evolutions of $A_{a}^{i}$ and $\E^{a}_{i}$ are decoupled. This is \emph{not} the case.  Because the electric shift vector $N^{a}{}^{i}_{j}$ is a `q-number field' that depends on $\E^{a}_{i}$, the `time' derivative of $A_{a}^{i}$ also depends on $\E^{a}_{i}$. Similarly, because the GGC Lie derivative depends on the connection $A_{a}^{i}$, the `time' derivative of $E^{a}_{i}$ also depends on $A_{a}^{i}$. 

2.  Interestingly, it turns out that the full set of equations on $\psyml$ automatically constitutes a symmetric hyperbolic system, making it directly useful to evolve  initial data using numerical or analytical approximation schemes (see, e.g., \cite{reula,shinkai}).

3.  In Section \ref{s1} we noted that the scalar and the vector constraints can be combined in a single spinorial equation (\ref{blending1}). When smeared with a spinorial `lapse-shift' field $\N^{A}{}_{B}$, it becomes
\be \label{blending3}
\int_{\Sigma}\,  \rmd^{3}x\,  \N^{A}{}_{D} \big(\E^{a}_{A}{}^{B}\,  \E^{b}_{B}{}^{C} \, F_{ab\, C}{}^{D}\big)  \simeq 0 \ee
As one might expect, this blended equation is equivalent to (\ref{blending2}), \emph{provided} the spinorial lapse-shift  $ \N^{A}{}_{D}$ is a `q-number field'  $\N^{A}{}_{D} = \N\, \epsilon^{A}{}_{D} + \sqrt{2} S^{a} {\underaccent{\tilde}{E}}_{\!a}{}^{A}{}_{D}$ constructed from $\N, S^{a}$ using the \emph{inverse} of the spinorial momentum variable.  Because of this equivalence, constraints  (\ref{blending3}) also constitute a first class system in Dirac's terminology. However, because of the the `q-number dependence'  of $N^{A}{}_{D}$ the Poisson-algebra is much more complicated.

\subsection{Recovering the Riemannian and Lorentzian geometrodynamics}
\label{s4.2}

In Section \ref{s4.1} we considered the complex phase space $\psymc$ in the gauge theory framework and discussed constraints and dynamics.  In this picture,  the familiar GR notion of metrics and extrinsic curvature are secondary, derived quantities and therefore not at forefront.  We will now use the dictionary of Section \ref{s2} to recover them. Specifically, we will show that restrictions to appropriate `real' sections $\psymr$ and $\psyml$ of (\ref{evo2})  that hold on full $\psymc$ yield the familiar geometrodynamical evolution equations for both Riemannian and Lorentzian GR.

Let us recall from Eq. (\ref{k1}) that the difference of the action $(\mathbf{D}_{a} -\D_{a}) \lambda_{i} $ of the derivative operator $\mathbf{D}$ compatible with $\E^{a}_{i}$ and the phase space variable $\D$ is encoded in a Lie-algebra valued 1-form $\pi_{a}^{i}$ as follows: $(\mathbf{D}_{a} -\D_{a})\,\lambda_{i} = \ez_{ij}{}^{k}\, \pi_{a}^{j} \lambda_{k}$. In the Riemannian sector $\pi_{a}^{i}$ is real while in the Lorentzian sector it is pure imaginary; $\pi_{a}^{i} = i \kub_{a}^{i}$, with $\kub_{a}^{i}$ real. Let us therefore set $\pi_{a}^{i} = \gamma \kub_{a}^{i}$ and set $\gamma =1$ to recover the Riemannian sector and $\gamma =i$ to recover the Lorentzian sector at the end.  The curvatures $F_{ab\,i}{}^{k} \equiv \ez_{ij}{}^{k}F_{ab}{}^{j}$ and $R_{ab\,i}{}^{k}$ of $\D$ and $\mathbf{D}$ are then related by:
\be \label{curvatures} 
\ez_{ij}{}^{k}F_{ab}{}^{j} = R_{ab\,i}{}^{k} - 2\gamma \ez_{ij}{}^{k}\, \textbf{D}_{[a} \kub_{b]}{}^{j} +2\gamma^{2} \kub_{[a}{}^{k} \kub_{b]\, i} \, .\ee
Let us restrict ourselves to the Gauss constraint surface   $\D_{a}\E^{a}_{i} =0$. It is easy to verify that this constraint is equivalent to the condition that the tensor field $q^{-\f{1}{2}}\,\kub_{a}{}^{i}\, \E^{b}_{i} \,q_{bc}$ is symmetric on $\Sigma$. Now, as we discussed in Section \ref{s2},  the (density weighted) orthonormal triad $\Eub^{a}_{i}$ is given by   $\Eub^{a}_{i} = \gamma \E^{a}_{i}$ in both Riemannian and Lorentzian sectors and the extrinsic curvature is given by $k_{ab} = q^{-\f{1}{2}}\, \Eub^{c}_{i} \kub_{(a}{}^{i}\, \,q_{b)c}$. Therefore, it follows from (\ref{curvatures}) that on the Gauss constraint surface, one has:
\be \label{con3} 
 \E^{b}_{i} F_{ab}{}^{i} =  q^{\f{1}{2}}\,\,  D_{b} \big(k_{a}{}^{b} - k\delta_{a}{}^{b}  \big)
\quad {\rm and}\quad 
\gamma^{2}\,\ez^{ij}{}_{k} \E^{a}_{i} \E^{b}_{j} \, F_{ab}{}^{k} = - q \mathcal{R} \, - \, \gamma^{2} q \big(k_{ab} k^{ab}\,- \, k^{2} \big)\, . 
\ee
On the Gauss constraint surface of $\psym$ let us express the Hamiltonian on $\psymc$ 
\be \label{HSN} H_{0,\Sv, \N}  =   \int_{\Sigma} \rmd^{3}x \, \big[ {\textstyle{\f{1}{2}}} \, \N\, \ez^{ij}{}_{k}\, \E^{a}_{i} \E^{b}_{j} \,F_{ab}{}^{k} \, +\, S^{a} \E^{b}_{k}\, F_{ab}{}^{k} \big] \ee
in geometrodynamical variables, using  (\ref{con3}), the  fact that $\gamma^{2} = \epsilon$, and the relation $ \P^{ab} = -\epsilon \,q^{\f{1}{2}}\, (k^{ab} - k^{ab})$ between the geometrodynamical momentum and the extrinsic curvature.  We have:
\ba  \label{Hrel} H_{0,\Sv, \N}  &=& \int_{\Sigma} \rmd^{3}x\,\Big[{\textstyle{\f{1}{2}}}\, N\big( - \epsilon\, q^{\f{1}{2}}\,  \mathcal{R}\, -\, q^{-\f{1}{2}}\, (\P^{ab}\P_{ab} - {\textstyle{\f{1}{2}}} \P^{2}) \big)\, -\epsilon \,S^{a}\, D_{b}\, \P^{b}{}_{a}\Big] \nonumber\\
&=& {\textstyle{\f{1}{2}}}\,\epsilon\,   \int_{\Sigma}\rmd^{3}x\,\big[ N \t{C} + S^{a}\t{C}_{a} \big]\,\, \equiv\,\,  {\textstyle{\f{1}{2}}}\, \epsilon\, H_{\Sv, N}  \ea 
where we have used for form (\ref{admcon}) of the geometrodynamical constraints $\t{C}$ and $\t{C}^{a}$, set $N = q^{\f{1}{2}}\, \N$, and denoted the geometrodynamical Hamiltonian by $ H_{\Sv, N}$. Thus,  the restriction of $ H_{0,\N, \Sv}$ to the Riemannian and Lorentzian sections yields the respective geometrodynamical Hamiltonians up to an overall factor of $\f{1}{2}\epsilon$. As discussed in the Remark below, this factor is exactly compensated by the  relation between the geometrodynamical Poisson bracket (\ref{pbadm}) and the gauge theory Poisson bracket (\ref{fpb3}) that follows from the dictionary. The pull-back of the symplectic structure to each of $\psymr$ and $\psyml$  is non-degenerate. Therefore, the restrictions of the Hamiltonian vector field $X_{H_{0,\N,\Sv}}$ to each of these `real' sections of $\psymc$ yields the Hamiltonian vector field of geometrodynamics for the two theories.

In terms of geometrodynamical variables, these vector fields have a complicated expression, 
just as one would expect from the fact that expressions of $q_{ab}, \P^{ab}$ in terms of $A_{a}^{i}, E^{a}_{i}$ involve non-polynomial  functions of $q_{ab}$. We have:
\be \label{HamSNadm} {X}_{\Sv,N}^{\rm geo}  = \dot{q}_{ab} \f{\delta}{\delta q_{ab}} + \dot{\P}^{ab} \f{\delta}{\delta \P^{ab}}\ee
with
\ba\label{admevo2}
\dot{q}_{ab} &=& \l_{\Sv}\, q_{ab}\, -\, 2\epsilon\, N q^{-\f{1}{2}} \big(q_{ac} q_{bd} -{\textstyle\f{1}{2}}  q_{ab} q_{cd})\, \P^{cd}\qquad {\rm and,}\nonumber\\
\dot{\P}^{ab} &=& \l_{\Sv}\, \P^{ab} \, + \,q^{\f{1}{2}}\big(q^{ac} q^{bd} - q^{ab} q^{cd} \big) D_{c}D_{d} N\, - \,q^{\f{1}{2}} N \big(q^{ac} q^{bd} - \,\,{\textstyle\f{1}{2}} q^{ab} q^{cd} \big) \,\mathcal{R}_{cd} \nonumber\\
&+& \epsilon\,  q^{-\f{1}{2}}  N\,\big(2\delta^{a}_{d} \delta^{b}_{n}q_{cm} - \delta^{a}_{m} \delta^{b}_{n}q_{cd} - {\textstyle\f{1}{2}} q^{ab} (q_{cm} q_{dn}  - {\textstyle\f{1}{2}} q_{cd} q_{mn}) \big) \P^{cd} \P^{mn}\, \ea
where, again, $\epsilon\! =\!1$ for the Riemannian signature and $\epsilon\! =\!-1$ for the Lorentzian. The contrast between the evolution equations (\ref{evo2}) and (\ref{admevo2}) is quite striking. It brings out the tremendous technical simplification and structural insights on `time evolution' that are made possible by putting the gauge theory framework at forefront, and using GGC Lie derivatives that preserve manifest gauge covariance. \emph{This is our main result.} As we will see in the next sub-section, the compact and geometric forms of the action of the Hamiltonian vector field on $\psymc$ greatly simplifies the task of calculating the constraint Lie algebra in both signatures.\\

\emph{Remarks:} \\
1. Note that the complete dictionary that leads to the geometrodynamical variables  \emph{also includes} $(\Lambda^{i}=0, S^{a}, \N) \to (S^{a}, N = q^{\f{1}{2}}\, \N)$.

2. The function $H_{\Sv, N}$ has a simple geometrical interpretation in the space-time picture. Consider a metric $g_{ab}$ on $\Sigma \times \mathbb{R}$, with Einstein tensor $G_{ab}$, that induces a positive definite metric $q_{ab}$ and extrinsic curvature $k_{ab}$ and a unit normal $n^{a}$ on a leaf $\Sigma$. Then,  as is well-known, the geometrodynamical Hamiltonian $H_{\Sv, N}$ is  given by $H_{\Sv, N} = 2\epsilon\, \int_{\Sigma} \rmd^{3}x\, G_{ab} (N\,n^{a}\, +\, S^{a}) n^{b}$, whence  the gauge theory Hamiltonian satisfies $H_{0,\Sv,\N} = \int_{\Sigma} \rmd^{3}x\, G_{ab} (N\,n^{a}\, +\, S^{a}) n^{b}$ on the Gauss constraint surface of  \emph{both} Riemannian and Lorentzian sectors. This is noteworthy first because $H_{\Sv,\N}$ was introduced just by appealing to simplicity and gauge invariance and second because its expression (\ref{HSN}) on $\psymc$ does not contain $\epsilon$.

3. On $\psymc$ we have a fixed Poisson bracket (\ref{fpb3}) and a fixed Hamiltonian (\ref{HSN}). \emph{There are no factors of $\epsilon$ that distinguish the Riemannian section from the Lorentzian in either.}  On the other hand, the form of the geometrodynamical Hamiltonian $H_{\Sv,N}$ changes as we pass from the Riemannian to the Lorentzian section because of the explicit  $\epsilon$ dependence. Still, restrictions to $\psymr$ and $\psyml$ of the Hamiltonian vector field $X_{H_{0,\Sv,\N}}$ of $H_{0,\Sv,\N}$ coincide with the geometrodynamical Hamiltonian vector fields on both sectors! This occurs because the relation between the Poisson brackets induced on $\psgeo$ by those on $\psymc$, noted in the remark at the end of Section \ref{s2}, just cancels out the factor of $2\epsilon$ relating the gauge theory $H_{0,\Sv,\N}$ and the geometrodynamical $H_{\Sv, N}$.  

 \subsection{Implications of the Gauge theory formulation } %Features 
 \label{s4.3}

The gauge theory formulation brings to the forefront several structures that are opaque in the metric formulation and also introduces significant technical simplifications.  In this section we will illustrate these features. The section is divided into three parts. In the first, we discuss (anti-)self-dual solutions that have featured in some of the recent literature on scattering amplitudes. This discussion will bring out the fact that  the evolution equations (\ref{evo1}) with the GGC Lie derivatives are natural generalizations to the full theory of the simple equations one encounters in these `half-flat' sectors. In the second we will show that the use of GGC Lie derivatives considerably simplifies the Poisson bracket between two scalar constraints. In the third we discuss two straightforward extensions of our results.

\subsubsection{The half flat sectors of GR}
\label{s4.3.1}

Self-dual and anti self-dual --or half flat-- solutions to Einstein's equations drew a great deal of attention in the 1970s and 1980s as they represent `exactly soluble' or `integrable' sectors of the theory \cite{newman,plebanski,penrose}.  However, the relation between these half-flat solutions and general solutions has remained elusive.  There is a revival of interest in this issue, reflected in the  recent investigations of the maximum helicity violating (MHV) processes in the scattering amplitude community (see in particular \cite{mason1}). We will now show that our geometrical formulation of `time evolution' in terms of GGC Lie derivatives serves to directly bring out the relation between dynamics of half flat solutions and general solutions.

Since we are interested in the `on-shell' structure,  let us begin by collecting the full set of Einstein's equations: the constraints (\ref{constraints})
\be \label{constraints2} \D_{a} \E^{a}_{i} =0; \quad  \E^{b}_{i} F_{ab}{}^{i} =0;\quad {\rm and} \quad \ez^{ij}{}_{k}\E^{a}_{i} \E^{b}_{j} F_{ab}{}^{k} =0 \, ; \ee
and the evolution equations (\ref{evo1}) with zero shift, restricted to the Gauss constraint surface:
\be \label{evo3} \dot{A}_{a}^{i}  = - \ez^{ij}{}_{k}\, \ggcl_{\Nv_{j}}  A_{a}^{k}\, \equiv\, - \ez^{ij}{}_{k} \,N^{b}_{j} F_{ba}{}^{k} \qquad {\rm and}\qquad
\dot{E}^{a}_{i}  =  {\textstyle{\f{1}{2}}}\, \ez_{i}{}^{jk}\, \ggcl_{\Nv_{j}}  E^{a}_{k}\, . \ee
(We have set $S^{a}=0$ because the main point can be made already in this case. Allowing non-zero shift is straightforward.) Recall that, on dynamical trajectories on the constraint surface, the curvature $F_{ab}{}^{i}$ of  $A_{a}^{i}$ turns out to be the self-dual part of the full Riemann curvature  (or, of the Weyl tensor since the Ricci tensor vanishes on solutions).  Therefore, we can recover the anti self-dual solutions, simply by setting $F_{ab}{}^{i} =0$. If we had changed the dictionary and set $A_{a}^{i} = \Gamma_{a}^{i} + \pi_{a}^{i}$ (respectively,  $A_{a}^{i} = \Gamma_{a}^{i} + i\pi_{a}^{i}$) this procedure would have given us the self-dual solutions in the Riemannian (respectively Lorentzian) sector.  Thus, the gauge theory framework is well-suited to investigate either type  --i.e. `half-flat' solutions. For definiteness let us use the dictionary given in Section \ref{s2} and discuss anti self-dual solutions. Also, for simplicity let us restrict ourselves to the Riemannian signature +,+,+,+  (just to avoid a digression associated with the choice of orientation in the Lorentzian signature where half flat solutions are complex, see \cite{ajs}).

Now, since $F_{ab}{}^{i}=0$, the vector and the scalar constraints are trivially satisfied, and $\dot{A}_{a}^{i}$ vanishes. Thus we are left only with the Gauss constraint, and the evolution equation for $\E^{a}_{i}$.  Since $A_{a}^{i}$ is pure gauge, we can  simplify the remaining equations by going to a gauge in which $A_{a}^{i}$ itself vanishes. This choice has three consequences: (i) The evolution equation on $A_{a}^{i}$ says that $A_{a}^{i}$ remains zero for all times; (ii) The covariant derivative $\D$ in the Gauss law reduces to the flat $\partial$ in that gauge; and, (iii) The GGC derivative in the evolution equation for $\E^{a}_{i}$ becomes the ordinary Lie derivative.  Thus, the remaining two equations assume an even simpler form
\be \partial_{a} \E^{a}_{i}=0 \qquad {\rm and} \qquad  \dot{\E}^{a}_{i} = {\textstyle{\f{1}{2}}}\,\ez_{i}{}^{jk}\, \l_{\Nv_{j}}
\E^{a}_{k} \ee
(Note that the first equation is well-defined because divergence of a vector density is independent of the choice of the derivative operator on the vector index.) Finally, let us fix a (nowhere vanishing) volume 3-form $\epsilon^{o}_{abc}$ that is annihilated by $\partial_a$ and denote its inverse by $\epsilon_{o}^{abc}$.  Such a choice always exists but is not unique; let us just choose one and then set the lapse to $\N  = {\textstyle{\f{1}{3!}}}  \epsilon_{o}^{abc} \,  \undertilde{\eta}_{abc}$. Then $\partial_{a} {\N} =0$. It is convenient to multiply both equations by $\N$ to get rid of the density weight of $\E^{a}_{i}$ and recast the equations entirely in terms of the electric shift vector fields $N^{a}_{i}$. We obtain:
\be \label{asd} \partial_{a} N^{a}_{i}=0 \qquad {\rm and} \qquad  \dot{N}^{a}_{i} = {\textstyle{\f{1}{2}}}\,\ez_{i}{}^{jk} \,\l_{\Nv_{j}}
N^{a}_{k} 
\ee
This is the \emph{full} content of Einstein's equations in the anti self-dual sector (at least locally). Thus, to obtain a solution, we only need to choose three vector fields $N^{a}_{i}$ that Lie-drag the pre-specified volume form, $\l_{\Nv_{i}} \epsilon^{o}_{abc} =0$, and evolve them via $\dot{N}^{a}_{i} = {\textstyle{\f{1}{2}}}\,\ez_{ij}{}^{k} \,\l_{\Nv_{j}} N^{a}_{k}$.  The first equation is trivially preserved under evolution.   Thus, \emph{the group of volume preserving diffeomorphisms is intimately intertwined with half-flat solutions to Einstein's equations.} (Recall that the self-dual YM equations are intertwined with area preserving diffeomorphisms \cite{sdym}.)

To specify the 4-metric defined by this dynamical trajectory in $\psym$, let us consider $M = \Sigma\times \mathbb{R}$, foliate it, and label the leaves by $t= {\rm const}$. Introduce a vector field $t^{a}$ on $M$ that is transverse to the leaves $\Sigma_{t}$ and has $t$ as its affine parameter; $t^{a}\partial_{a}t = 1$. Then, the 4-metric on $M$ that solves the anti self-dual Einstein's equation is given by
\be  g^{ab} = (\hat{q})^{\f{1}{2}}\, \N \big(\hat{q}^{ab} + t^{a} t^{b}\big) \quad {\rm where} \quad
\hat{q}^{ab} = N^{a}_{i} N^{b}_{j} \qz^{ij}  \ee  
Locally, \emph{every} anti self-dual solution has this form, i.e., it results from solutions to (\ref{asd}). 

This final result has been in the literature for quite some time \cite{ajs}. At the time, the simplicity  and the geometric nature of the evolution equation in (\ref{asd}) came as a surprise and its `origin' has continued to be unclear. In particular, it was puzzling to see the \emph{time}-evolution of a triad vector expressed only in terms of  its Lie derivatives along \emph{spatial} vector fields. The common viewpoint was that this intriguing feature is an artifact of the restriction to half flat solutions. Our formulation of dynamics on the gauge theory phase space of  Section \ref{s4.1} shows that this view is incorrect: The intermingling between `time' and `space' evolution also occurs in \emph{full} GR.  Essentially the only simplification in the evolution of the triad in the half flat sector is that the GGC Lie derivative $\ggcl_{\Nv_{i}}$ in the general case (\ref{evo1}) is now replaced by an ordinary Lie derivative. Technically, this is of course an enormous simplification because the evolution of the triad is then decoupled from the evolution of the connection. But conceptually, the phenomenon descends from full GR. (In fact,  if we had not gauge fixed  $A_{a}^{i}$ to  zero, we would have the GGC Lie derivative $\ggcl_{\Nv_{j}}$ in place of $\Nv_{j}$ in the evolution equation (\ref{asd})  of the triad also in the anti self-dual sector.)  

Conversely, the evolution equation in full GR can be regarded as the most natural  --i.e., gauge covariant--  generalization of that in the `integrable' half-flat sector, in that the ordinary Lie derivative is simply replaced by a GGC Lie derivative. Since the anti self-dual sector is naturally embedded in the full phase space $\psym$, the framework of Section \ref{s4.1} is well suited  for investigating  the  tree-level MHV processes by  investigating the dynamics of self-dual perturbations on anti self-dual backgrounds \cite{mason1} from a classical perspective using Hamiltonian methods. 

\subsubsection{The Constraint algebra}
\label{s4.3.2}

Let us now turn to the Poisson algebra of constraints in the full theory. Since  the canonical transformations generated by the vector \emph{as well as the scalar}  constraint have a simple geometric interpretation in terms of GGC Lie derivatives, we will find that the computation of this algebra on $\psymc$ is significantly simpler than that in geometrodynamics.

Let us begin with the Gauss constraint 
\be \G_{\Lambda} (A,\E) := \int_{\Sigma} \rmd^{3} x\, \Lambda^{i}\, (\D_{a}\E^{a}_{i})\,  \simeq  0\ee
As in any gauge theory, we have $\{ \G_{\Lambda_{1}},\,  \G_{\Lambda_{2}} \} \, =\, - \G_{\Lambda_{3}}$, \, where $\Lambda_{3}^{i} =  \ez^{i}{}_{jk} \Lambda_{1}^{j} \Lambda_{2}^{k}$. 
The calculation of the Poisson brackets between the Gauss constraint and the vector 
\be \V_{\Sv} (A, \E) :=  \int_{\Sigma}\, \rmd^{3}x \,S^{a} \E^{b}_{i}\, F_{ab}{}^{i}\, \simeq 0; \ee
 and the scalar 
 \be \tilde{\mathcal{S}}_{\N}(A, \E) :=  {\textstyle{\f{1}{2}}}\int_{\Sigma}\, \rmd^{3}x \,\N \E^{a}_{i}\, \E^{b}_{j}\, F_{ab}{}^{k} \ez^{ij}{}_{k}\, \rmd^{3}x \equiv  {\textstyle{\f{1}{2}}}\int_{\Sigma}\, \rmd^{3}x \,N^{a}{}^{j}{}_{k} \E^{b}_{j}\, F_{ab}{}^{k} \, \simeq 0
\ee
constraints is trivialized by the fact that  $\V_{\Sv} (A, \E)$ and $\tilde{\mathcal{S}}_{\N}$ are gauge invariant. Therefore we have
\be \label{pb1} \{ \G_{\Lambda_{1}},\,  \G_{\Lambda_{2}} \} \, =\,  - \G_{\Lambda_{3}}, \quad   \{\G_{\Lambda_{1}},\, \V_{\Sv}\}  =0, \quad{\rm and} \quad  \{\G_{\Lambda_{1}},\, \tilde{\mathcal{S}}_{\N} \} =0 \ee
everywhere on the phase space, where $\Lambda_{3}$ is defined above. Next, let us consider the Poisson brackets between two vector constraints.  We have
\be \{ \V_{\vec T}\,, \,\,\, \V_{\Sv}\} \,=\, X_{H_{\Sv}} \,\, {\Large\vins}\,\, \mathbf{d} H_{\vec T}\,\ee
where the right side denotes the contraction of the Hamiltonian vector field $X_{H_{\Sv}}$ with the exterior derivative of $H_{\vec T}$ on the infinite dimensional space $\psymc$. Using the explicit form (\ref{hamS}) of $X_{H_{\Sv}}$, we obtain
 \ba \label{pb2} \{ \V_{\vec T} \,,\,\,\, \V_{\Sv}\} \,&=& \int_{\Sigma}\,\rmd^{3}x\,  \big[T^{a}\,  \gcl_{\Sv}\, (\E^{b}_{i}\, F_{ab}{}^{i} ) 
 - (T^{a} S^{b} F_{ab}{}^{i})\, \D_{c} \E^{c}_{i}  \big]\nonumber\\
 &=& - \int_{\Sigma} \,\rmd^{3}x\, \big[ (\l_{\Sv}\,T^{a})\,   \E^{b}_{i}\, F_{ab}{}^{i} 
 + (T^{a} S^{b} F_{ab}{}^{i})\, \D_{c} \E^{c}_{i} \big]\, \nonumber\\
 &=:&\, (\V_{U} - \G_{\Lambda}) \ea
 where
 \be U^{a} =  \l_{\vec T} S^{a}   \qquad {\rm and} \qquad  \Lambda^{i} = T^{a} S^{b} F_{ab}{}^{i} \, ,\ee
 just as one would expect from the expression (\ref{com2}) of the action of the GC commutator of two vector fields on an object with internal index. In the last step of (\ref{pb2}) we have used the fact that, since $\E^{a}_{i}F_{ab}{}^{i}$ has no free internal indices, the action of the GC Lie derivative is the same as that of ordinary Lie derivative, and then performed an integration by parts. Next consider the Poisson bracket between the vector and the scalar constraint. Repeating the steps that led us to (\ref{pb2}),  we immediately obtain
 \be \{ \V_{\Sv} \,,\,\,\, \S_{\N}\} \,=  (\S_{\underaccent{\tilde}{L}} \,-\, \G_{\Lambda}) \quad {\rm with} \quad
 \S_{\underaccent{\tilde}{L}} = \l_{\Sv} \N  \,\,\, {\rm and}\,\,\, \Lambda^{i} = 2 S^{a} N^{b}{}^{i}{}_{k}\, F_{ab}{}^{k} \, \ee
 again, just as one would expect from (\ref{com2}), since the scalar  constraint has the same form as the vector constraint with an electric shift $N^{a}_{i}$.
 
Finally, let us consider the Poisson bracket between two scalar constraints whose calculation is very long in geometrodynamics (see, e.g., pp 52-54 of \cite{ttbook}).  %where there are still missing steps each of which, to quote the author, would  require ``at least one Din A4  page of calculations'').  
We will again use the identity 
\be  \{ \S_{\M} \,,\,\,\, \S_{\N}\}\,=\, X_{H_{\N}} \,\, {\Large\vins}\,\, \mathbf{d} H_{\M} \, 
\equiv\, {\textstyle{\f{1}{2}}} \big(X_{H_{\N}} \,\, {\Large\vins}\,\, \mathbf{d} H_{\M}  - X_{H_{\M}} \,\, {\Large\vins}\,\, \mathbf{d} H_{\N}\big) \, \, 
\ee
and  the expression of the Hamiltonian vector field of  $\S_{N} \equiv H_{0,0,\N}$ from (\ref{hamN}). Since  the Poisson bracket is manifestly antisymmetric in $\N$ and $\M$, terms proportional to the Gauss constraint in the expression (\ref{hamN}) of the Hamiltonian vector field cancel because $\N$ and $\M$ sit `outside' the derivative operator in this term. Therefore, we only have to consider the GGC Lie derivative terms in (\ref{hamN}).  Hence we have:
 \ba \label{pb4}  \{ \S_{\M} \,,\,\,\, \S_{\N}\} \, &=&  {\textstyle{\f{1}{4}}}\, \int_{\Sigma} \rmd^{3}x \, \M\, \ez^{ij}{}_{k}\, \big(\dot{\E}^{a}_{i}\, \E^{b}_{j}\, F_{ab}{}^{k} + \E^{a}_{i}\dot{\E}^{b}_{j}\, F_{ab}{}^{k} +
 \E^{a}_{i}\,\E^{b}_{j}\, \dot{F}_{ab}{}^{k} \big)\,\, - \M \leftrightarrow \N \nonumber\\
 &=&  {\textstyle{\f{1}{4}}}\, \int_{\Sigma} \rmd^{3}x \, \M\, \ez^{ij}{}_{k}\, \Big[ \ez_{i}{}^{mn} \,(\ggcl_{\Nv_{m}}\,  \E^{a}_{n})\, \E^{b}_{j}\, F_{ab}{}^{k} \,\,-\,\, \E^{a}_{i}\,\E^{b}_{j}\, \ez^{km}{}_{n}\, \ggcl_{\Nv^{m}} F_{ab}{}^{n} \Big]  - \M \leftrightarrow \N  \nonumber\\
&=& {\textstyle{\f{1}{4}}}\, \int_{\Sigma} \rmd^{3}x \, \M\, \big[ (\ggcl_{\N_{j}} \E^{a}_{k} -\ggcl_{\N_{k}} \E^{a}_{j})\, \E^{b\, j}\, F_{ab}{}^{k}\,\,+\,\, \E^{a}_{k} \E^{b}_{j}\, (\ggcl_{\Nv^{j}} F_{ab}{}^{k} - \ggcl_{\Nv^{k}} F_{ab}{}^{j} ) \Big]- \M \leftrightarrow \N  \nonumber\\ 
 &=&  {\textstyle{\f{1}{4}}}\, \int_{\Sigma} \rmd^{3}x \, \Big[ 2\M\,\ggcl_{\Nv_{j}} (\E^{a}_{k}\, F_{ab}{}^{k})\, \E^{b\,j}\,\, +\,\, 2\M \E^{a\, (k} F_{ab}{}^{j)}\, (\ggcl_{\Nv_{k}} \E^{b}_{j})\, \Big] - \M \leftrightarrow \N  \nonumber\\
 &=&   {\textstyle{\f{1}{2}}}\, \int_{\Sigma} \rmd^{3}x \,\big( \ggcl_{\Nv^{j}} \M^{a}_{j}\, -\, \ggcl_{\Mv^{j}} \N^{a}_{j}\big)\, \E^{b}_{k}\,F_{ab}{}^{k}  \nonumber\\
 &=& \V_{\vec{V}}\qquad {\rm with}\qquad V^{a} ={\textstyle{\f{1}{2}}} \,\big(\ggcl_{\Nv^{j}} \M^{b}_{j}\, -\, \ggcl_{\Mv^{j}} \N^{b}_{j}\big)\,\, \equiv\,\, \ggcl_{\Nv^{j}} \M^{b}_{j} \, . 
 \ea
 Here, in the last but one  (i.e., 5th) step, we have expanded out $\ggcl_{\Nv_{k}} \E^{b}_{j}$, substituted $N^{a}_{k}= \N \E^{a}_{k}$ and used the antisymmetry in $\N$ and $\M$ to show that  the term $\big(\M\E^{a\, (k} F_{ab}{}^{j)}\, (\ggcl_{\Nv_{k}} \E^{b}_{j}) -  \N \leftrightarrow \M\big)$  vanishes. Thus, as expected, the Poisson bracket between two scalar constraints is a vector constraint. However, the calculation is conceptually stream-lined and \emph{much} shorter than the one in geometrodynamics  because we could use properties of the GGC Lie derivative $\ggcl_{\Nv_{j}}$ and integrate by parts, just as in the case of the Poisson bracket between two vector constraints. Note that, because we recast the scalar constraint as a vector constraint smeared with the electric shift, this calculation of the Poisson bracket is quite similar to the simpler calculation of the Poisson bracket between two vector constraints. Indeed,  the smearing vector field $V^{a} = \ggcl_{\Nv^{j}} \M^{b}_{j} $ in the last but one step mirrors the  smearing vector field $U^{a} = \l_{\vec T} S^{a}$ in the expression (\ref{fpb2}) of the Poisson bracket between two vector constraints. (The only difference is that each of the vector fields $\Nv_{j}$ and $\Mv_{j}$ now carries an internal index; it is summed over as it must be for the result to yield a smearing field for the vector constraint.) This is another concrete and deeper consequence of the fact that the `time evolution' is now brought to the same footing as the `space' evolution. We presented every step in this calculation to bring out how the new conceptual input enormously shortens the calculation.
 
  Finally, one can expand out the GGC Lie derivative in the expression of $V^{a}$ to obtain the more familiar but less geometric expression
 \ba \label{VNM}  V^{a} &=& \ggcl_{\Nv^{j}} \M^{a}_{j} \nonumber\\
 &=& \E^{a}_{i} \E^{b}_{j} \,\qz^{ij}\, \big(\N \partial_{b} \M - \M \partial_{b} \N\big)  \equiv 
 \epsilon\, q^{ab} \big(N \partial_{b} M - M \partial_{b} N\big)  \ea
where, in the last step, we have used the dictionary to set $N = q^{-\f{1}{2}}\ \N$ and $q^{ab} =\epsilon\, q^{-1}\, \E^{a}_{i}\, \E^{b}_{j}\,\qz^{ab}$. Thus, the structure function $\epsilon\, q^{ab} \big(N \partial_{b} M - M \partial_{b} N\big)$  that appears in geometro\-dynamics has a simple geometric meaning of a GGC Lie derivative between electric shifts on the gauge theory phase space.\smallskip
  
 Let us conclude by collecting all the Poisson brackets:
 \ba  &\{ \G_{\Lambda_{1}},\,  \G_{\Lambda_{2}} \} \, =\, - \G_{\Lambda_{3}}, \qquad  
 \{\G_{\Lambda_{1}},\, \V_{\Sv} \}  &=0, \qquad  \{\G_{\Lambda_{1}},\, \tilde{\mathcal{S}}_{\N} \} =0,  \qquad{\rm and}  \nonumber\\ 
 &\{ \V_{\vec T} \,,\,\,\, \V_{\Sv}\} =  (\V_{\vec{U}} - \G_{\Lambda}), \qquad   \{ \V_{\Sv} \,,\,\,\, \S_{\N}\} \,&=  (\S_{\underaccent{\tilde}{L}} \,-\, \G_{\bar\Lambda}),\qquad  \{ \S_{\M} \,,\,\,\, \S_{\N}\} = \V_{ \vec{V}}  \nonumber \ea
 where,
 \ba  &\Lambda_{3}^{i}  =  \ez^{i}_{jk}\, \Lambda_{1}^{j}\, \Lambda_{2}^{k}, \qquad
 &\Lambda^{i} = \, T^{a} S^{b} F_{ab}{}^{i}, \qquad \bar\Lambda^{i} = S^{a}N^{b}{}^{i}{}_{k}\, F_{ab}{}^{k},\nonumber\\
  &U^{a} = \l_{\vec T} S^{a}, \qquad  &V^{a} =\ggcl_{\Nv^{j}} \M^{b}_{j}, \qquad  {\underaccent{\tilde}{L}} = \l_{\Sv} \N\, .
 \ea
 %
% \be \Lambda_{3}^{i}  =  \ez^{i}_{jk}\, \Lambda_{1}^{j}\, \Lambda_{2}^{k}, \quad  U^{a} = \l_{\Sv} T^{a}, \quad  \M = \l_{\Sv} \N, \quad \bar\Lambda^{i} = N^{a}{}^{i}{}_{k}\,S^{b} F_{ab}{}^{k}, \quad V^{a}=\ggcl_{\Nv^{j}} \M^{b}_{j}
%\ee
% 
\subsubsection{Extensions}
\label{s4.3.3}

In the main body of the paper, for simplicity, we assumed that the 3-manifold $\Sigma$ is compact.  However, all our results go through also in the asymptotically flat case, once suitable boundary conditions are imposed on the canonical variables and the lapse and shift fields, and surface terms are included in the expression of the Hamiltonian. For definiteness, we will consider the Lorentzian case and use the standard asymptotic boundary conditions on the the phase space variables. Specifically, the real triad $\Eub^{a}_{i}$ and the real field $\kub_{a}^{j}$  --that determines the extrinsic curvature on solutions--  have to be subject to the standard asymptotically flat boundary conditions spelled out in, e.g., \cite{newvar,aabook,ttbook}. To obtain the Hamiltonian that generates asymptotic translations, one considers lapse-shift pairs $\N, S^{a}$ such that $\N\to (q^{o})^{-\f{1}{2}} + O(\f{1}{r})$  --where $q^{o}_{ab}$ is the asymptotic,  flat metric and $r$ its radial coordinate--  and $S^{a}$ tends to a unit space-translation $S^{a}_{o}$ of $q^{o}_{ab}$ as $\f{1}{r}$. Then, the function
\ba \label{ham3} H_{\Sv, \N}(A,E) &=& {\textstyle{\f{1}{2}}}\, \int_{\Sigma} \rmd^{3}x \,\big[ \ez^{ij}{}_{k}\, \N\,\E^{a}_{i} E^{b}_{j} F_{ab}{}^{k} \, +\, 2S^{a} \E^{b}_{k} F_{ab}{}^{k} \big] \nonumber\\
&-&  \oint_{\partial\Sigma} \rmd^{2} S_{a} \big[\ez^{ij}{}_{k} \,\N\, \E^{a}_{i} \E^{b}_{j} A_{b}^{k} 
+2 S^{a} \E^{b}_{k}\, F_{ab}{}^{k} \big]\, 
 \ea
is differentiable on the asymptotically flat phase space and serves as the Hamiltonian generating the asymptotic translation defined by $(\N, S^a)$. Both the volume and surface integrals are real on the Lorentzian section. The boundary term is proportional to the ADM 4-momentum, and exactly equals the ADM 4-momentum if one rescales numerical factors in our dictionary to get exact agreement with the ADM symplectic structure, as indicated at the end of Section \ref{s4.2}. The infinitesimal canonical transformations it generates are the same as in Section \ref{s4.1}, so that the Hamiltonian vector field is again given by (\ref{hamSN}).  Therefore, all the results of Sections \ref{s4.1} and \ref{s4.2} go through also in this case; time evolution emerges as GGC Lie derivatives along \emph{spatial} electric shift fields. 

Finally, the  framework can also be naturally extended to include a scalar field $\phi$ (with a potential $V(\phi)$) that sources the gravitational field. Now the phase space $\psymc$ is extended to include a pair of real, canonically conjugate fields fields $(\phi, \t\pi)$ on $\Sigma$, and the Hamiltonian that governs the dynamics of the combined system is given by
\ba H_{0,\Sv, \N} (A,\E;\,\phi, \t\pi) =  &{\textstyle{\f{1}{2}}}& \,\int_{\Sigma} \rmd^{3}x\, \big[ \ez^{ij}{}_{k}\, \N\,\E^{a}_{i} \E^{b}_{j} F_{ab}{}^{k} \, +\, 2 S^{a} \E^{b}_{k} F_{ab}{}^{k} \big]\nonumber\\
&-& {\textstyle{\f{1}{2}}} \,\int_{\Sigma} \rmd^{3}x\, \big[ \N (\t\pi^{2} +V(\phi) q) - \N^{-1}  (\ggcl_{\Nv_{i}} \phi ) (\ggcl_{\Nv_{j}} \phi )\qz^{ij}   
\,+\, 2\t\pi\, \l_{\Sv} \phi\big] \nonumber\\  
&-&  \oint_{\partial\Sigma} \rmd^{2} S_{a} \big[\ez^{ij}{}_{k} \,\N\, \E^{a}_{i} \E^{b}_{j} A_{b}^{k} 
+ 2S^{a} \E^{b}_{k}\, F_{ab}{}^{k} \big]\, .
\ea
Note that the scalar field contribution is also expressed only in terms of canonical variables and their GGC and ordinary Lie derivatives (and of course the lapse $\N$ and the shift $S^{a}$). The Hamiltonian continues to be a low order polynomial in the basic canonical variables and therefore the equations of motion also share this property.

\section{Discussion}
\label{s5}

We will first summarize the main results and then make a number of remarks to put them in a broader perspective.

The point of departure of our investigation is the old observation that GR can be recovered from a background independent gauge theory that, to begin with, makes no reference to a metric or its curvature. While our discussion was based on the Hamiltonian perspective of \cite{newvar}, the observation also holds from a 4-dimensional space-time perspective \cite{cdj,cdjm}.  In this framework, the gauge theory concepts --particularly $SU(2)$ connections, their curvature and holonomies--  are at forefront. The emphasis is on fields with internal indices. Indeed,  the theory does not even need a specific derivative operator on tensor indices! The (pseudo)Riemannian metric, its causal structure,  derivative operator and curvature, are of course ubiquitous in GR and play a central role in classical physics. However, from the gauge theory perspective they are all `emergent' rather than primary notions. This perspective lies at the heart of LQG: The viewpoint is that the gauge theory framework --together with the associated rich set of techniques, particularly `Wilson loops'--  offers a better starting point for a background independent, non-perturbative  approach to quantum gravity  (see, e.g., \cite{alrev,ttbook,30years,crbook}). 

The LQG community is therefore very familiar with the `connection-dynamics' framework used in this paper. However, as noted in Section \ref{s1} one typically arrives at it starting from the metric and associated structures, whence the gauge theory perspective is often `diluted'. For example, one often uses the Poisson brackets $\{A_{a}^{i}(x) ,\, \E^{b}_{j} (y)\} = \delta_{a}^{b}\, \delta^{i}_{j}\, \delta^{3}(x,y)$ in the Riemannian theory and $\{A_{a}^{i}(x) ,\, \E^{b}_{j} (y)\} = i\, \delta_{a}^{b}\, \delta^{i}_{j}\, \delta^{3}(x,y)$ in the Lorentzian, and the expression of constraints often contain the signature dependent factors. In this paper, by contrast, both sectors arose as two `real sections' of the fixed gauge theory phase space $\psymc$.  Poisson brackets, constraints, and equations of motion  were specified once and for all on full $\psymc$ without any signature dependent factors. Yet, restriction to  the Riemannian `real section' yielded the constraints and equations of motion of Riemannian geometrodynamics, and the restriction to  the Lorentzian `real section' yielded the  constraints and equations of motion of  Lorentzian geometrodynamics. 

The more significant new element is that the `purer' gauge theory perspective with its emphasis on gauge covariance led us to a pleasing `blending' of the vector and the scalar constraints and `time' and `space' evolutions. Let us summarize how this rather unexpected unification came about. To begin with,  the classical Hamiltonian theory has 7 first class constraints.  The canonical transformations generated by six of them --the Gauss and the vector constraints-- have a natural geometrical interpretation in terms of the Lie algebra generated by infinitesimal gauge rotations and infinitesimal diffeomorphisms generated by vector fields on the 3-manifold $\Sigma$. However, the crux of dynamics lies in the seventh, scalar constraint. While canonical transformation it generates is also algebraically simple in the gauge theory formulation that has been used in LQG, its `structural' meaning had remained obscure. Our principal result is that:  (i)  this canonical transformation also has a simple geometrical meaning as a \emph{generalized gauge covariant {\rm (GGC)} Lie derivative}; (ii)  the Lie derivative in question is along \emph{spatial}  `electric shifts', whence `time evolution' is cast as a suitable `space evolution'; and (iii) the GGC Lie derivatives form an infinite dimensional Lie algebra  just as the gauge covariant (GC) Lie derivatives do, albeit it is a much larger, graded Lie algebra (see below). This interpretation of the dynamics is rooted in the gauge theory perspective, particularly the requirement of gauge covariance on geometrical operations such as Lie derivatives. This coming together of the two constraints  and their actions is perhaps as clear a manifestation of the underlying 4-dimensional covariance of the theory as we can hope for, within the realm of   Hamiltonian methods. \emph{Thus use the of electric shift leads to a conceptual shift in the description of dynamics in the Hamiltonian perspective.} Now, in the LQG literature, the connection-dynamics formulation has been extended to include supergravity \cite{susy1,susy2}. It would be interesting to investigate if the use of GGC Lie derivatives and electric-shifts also simplifies the dynamics there. 

The interpretation provides a new perspective on the structure of Einstein's equations. We presented two examples. First, we now have a unified picture of the smeared scalar and vector constraints --the former has the same form as the latter but the smearing field is a `q-number electric shift' $N^{a}_{i}$ rather than a c-number shift $S^{a}$. Consequently the constraint algebra is streamlined --it uses essentially the same operations for both vector and scalar constraints, resulting in a clear underlying unity in the nature of the `shift vector' $\l_{S} T^{a}$ that appears in the Poisson bracket of two vector constraints smeared with shifts  $S^{a}$ and $T^{a}$,  and the shift vector $\ggcl_{\Nv_{j}} M^{a}_{j}$  that appears in the Poisson bracket of two scalar constraints smeared with electric shifts $N^{a}_{j}$ and $M^{a}_{j}$. This result also illuminates `origin' of the $\epsilon$-dependent structure \emph{functions} $\epsilon\,q^{ab} (ND_{a} M - M D_{a} N)$ in the Poisson algebra of geometrodynamics: they arise simply because one expands out $\ggcl_{\Nv_{j}} M^{a}_{j}$. The reason why we have a structure function  --rather than a structure constant-- is simply that the scalar constraints are smeared with `electric shifts' $N^{a}_{j} = \N\E^{a}_{j}$  that are `q-number' vector fields, while the vector constraints are smeared with c-number vector fields $S^{a}$. 

Our second illustration used the `integrable' half-flat sectors of GR \cite{newman,plebanski,penrose}. That Einstein's equations assume the extremely simple form (\ref{asd}) in the half-flat sector, and are intertwined with the group of volume preserving diffeomorphisms, has been known for quite some time \cite{ajs}.  Our recasting the scalar constraint of full GR as a vector constraint with an `electric shift'  brought out the precise sense in which these features descend directly from full GR: in the half flat sector, the gauge covariant $\D$ can be replaced by a flat $\partial$  whence the GGC Lie derivative between the three electric lapse fields $N^{a}_{j}$ becomes ordinary Lie derivative. Thus, the full GR equations can be regarded as a natural gauge covariant generalization of those in the much simpler, half-flat sector. Now, it has been known for some time that the half-flat solutions to Einstein's equations are in 1-1 correspondence with hyperk\"ahler manifolds \cite{robinson}. It would be interesting to see if the notion of GGC Lie derivatives also sheds direct light on the `origin' of this rich structure. From the perspective of our gauge theory reformulation, can dynamics of full GR also be regarded as encoding a natural `gauge covariant extension' of hyperk\"ahler structures in the half flat sector?

Another set of mathematical issues  that may lead to new insights and applications arises from the notion of a GGC Lie derivative. The usual Lie derivative of two vector fields is of course just another vector field, and vector fields form a Lie algebra under this operation. By contrast, the `electric shift' $N^{a}{}_{i}$  is a vector field that takes values in $su(2)$. Therefore,  the GGC Lie derivative $\ggcl_{\Nv_{j}} M^{a}_{k} =: P^{a}{}_{jk}$ carries two internal indices rather than one.%
\footnote{It may be tempting to use the structure functions $\ez_{i}{}^{jk}$ to define the Lie derivative as  a vector field $\ez_{i}{}^{jk}\, \ggcl_{\Nv_{j}} M^{a}_{k}$ with a single internal index. However, this strategy does not work --this operation is not even antisymmetric in $N^{a}{}_{i}$ and $M^{a}{}_{i}$!}  
Consequently, the calculation of commutators leads one to consider GGC Lie derivatives by  vector fields $N^{a}{}_{i_{1} \ldots i_{n}}{}^{j_{1}\ldots j_{m}}$  with an arbitrary number of internal indices.  These fields have the structure of a graded vector space.  As Appendix A shows,  GGC Lie derivative they generate, together with local gauge transformations, form a graded Lie algebra.  Therefore, one would expect that there is a rich underlying mathematical structure with an infinite dimensional group whose action has a geometrical interpretation.  Let us make a small detour to explain this point.  As we saw in Section \ref{s3.1}, the gauge covariant (GC)  Lie derivatives generated by ordinary vector fields do not close under commutator. However, as we recall in Appendix A.2, together with local gauge transformations they do form an infinite dimensional Lie algebra. This is the Lie algebra of the semi-direct product of the group of local gauge transformations with the group of diffeomorphisms on $\Sigma$. Furthermore, we can realize this action geometrically as the group of structure preserving diffeomorphisms on  $su(2)$ bundles over $\Sigma$ \cite{jackiw}. Is there an analogous interpretation of the action of GGC Lie derivatives? If so, what is the bundle? Do we need to have the infinite dimensional, graded vector space of all tensors only with internal indices as fibers over $\Sigma$? Perhaps one could proceed in the spirit of the Gel'fand theory in which geometrical notions are recovered from suitable algebraic ones --e.g. the manifold is recovered knowing only its ring of regular functions; regular vector fields from derivations on the given ring; regular forms from linear mappings from the space of regular vector fields to regular functions; etc.  We introduced  the GGC Lie derivative in terms of its action on the graded algebra of fields with  internal indices, but they could perhaps be realized as geometric operations (say diffeomorphisms) on a suitable (possibly infinite dimensional) manifold. Can one reconstruct this manifold, say along the lines of the Gel'fand theory? These issues are interesting in themselves from a mathematical physics perspective. In addition, their resolution may lead to a deeper understanding of new infinite dimensional groups underlying GR which may, for example, very significantly broaden the notion of symmetries and associated conserved charges.\medskip

We will conclude with three remarks that elucidate different aspects of the gauge theory formulation of gravitational dynamics presented in this paper. \smallskip

1. The \emph{Belinskii Khalatnikov Lifshitz conjecture \cite{bkl}:} In the gauge theory phase space, we saw that `time' derivatives of the basic canonical pair $(A_{a}^{i}, \E^{b}_{j})$ can be represented by GGC Lie derivatives along \emph{spatial} directions. At first sight, this result may seem to negate the BKL conjecture which posits that time derivatives dominate over space derivatives near space-like singularities. However, there is no tension what so ever: We showed in Section \ref{s4.2} that the evolution equations for  $(A_{a}^{i}, \E^{b}_{j})$ imply those for the geometrodynamical variables $(q_{ab}, k_{ab})$  --the 3-metric and the extrinsic curvature-- that have been used to display the BKL behavior in various contexts (see, e.g., \cite{bkl-berger}).  The point is that the BKL conjecture refers to derivatives of the metric while, when translated to the geometrodynamical variables,  the GGC Lie derivative involves also the extrinsic curvature through the connection $A_{a}^{i}$ that features in the GC derivative operator $\D$. 

In fact there is a useful formulation of the  BKL conjecture that not only uses the gauge theory perspective of this paper but, in a sense, takes it to the extreme \cite{ahs}.  It leads to new insights into space-like singularities of GR, beyond the BKL conjecture. The key step is to  eliminate the vector index on the connection $A_{a}^{i}$ by contracting it with the triad to obtain a density $\t{A}^{i}{}_{j} := A_{a}^{i}\Eub^{a}_{j} $, and recast the constraints and evolution equations of full GR in terms of fields with \emph{only} internal indices.  Somewhat surprisingly,  it is possible to write the constraints and evolution equations using just these fields. Furthermore, in examples that have been studied in the cosmological context, these fields remain \emph{finite} at the singularity because the density weighted contravariant triad $\Eub^{a}_{i}$ goes to zero there. (So, these fields with only internal indices serve the role of the `Hubble normalized' tensor fields used in the standard, geometrodynamical formulation of the BKL conjecture, which admit well-defined limits at cosmological singularities.)  Therefore, one can adopt the following strategy to evolve initial data. Begin with  smooth data $(A_{a}^{i}, \E^{a}_{i})$ in the Lorentzian section $\psyml$, satisfying the constraints, on some initial slice and construct from it $\t{A}^{i}{}_{j}$. Then evolve it. In various examples, the evolution does not break down at the singularity, although the curvature of the space-time metric diverges there. Therefore one can `evolve' the (density weighted) fields with only internal indices across the singularity. Once one arrives on the `other side', one can attempt to reconstruct the metric because the evolution equation for $\Eub^{a}_{i}$ features only those quantities that remain well-behaved even at the time when the metric curvature diverges. (For details, see \cite{ahs}.)

Thus, because the equations in the gauge theory formulation do not break down even when the covariant metric $q_{ab}$ becomes  degenerate (or even vanishes, as at the big bang), it is possible to provide a specific time evolution across at least some singularities. This unforeseen and exciting possibility arises only because the standard geometrodynamical \emph{tensor} fields  --the metric $q_{ab}$, its curvature $\mathcal{R}_{ab}$, and the extrinsic curvature $k_{ab}$-- never feature in any of the equations. They are secondary, to be reconstructed as `derived quantities' starting from the gauge theory primary fields. Of course from the physical perspective of GR , the singularity is not resolved and physical quantities do diverge there.  Nonetheless, the procedure offers a natural avenue to continue the evolution across the singularity. A systematic investigation of the class of singularities that can be so `transcended'  has not been undertaken. The perspective on dynamics presented in this paper is likely to simplify that analysis. Is there a wide class of space-like singularities across which one can evolve using the gauge theory framework? 
\footnote{Note that, to make contact with general relativity --i.e. to reconstruct the metric-- the procedure requires one to first specify the geometrodynamical data on a complete Cauchy surface, convert it to the connection variables with only internal indices, and then evolve.  So, while the procedure is well-adapted to space-like singularities, it is not applicable if there is a time-like or null singularity that prevents one from completing the first step.}

2. \emph{Gauge theory/gravity correspondence:} This paper --as well as most of LQG-- is based on a formulation of GR as a background independent  \emph{gauge theory}.  A gauge theory/gravity correspondence also features, prominently, in string theory through the AdS/CFT proposal. However, there are some deep differences. Here, the correspondence uses a \emph{background independent} gauge theory in the \emph{bulk} which is then shown to be equivalent to GR  also in the bulk, with or without a cosmological constant  $\Lambda$. (In  this paper we focused on $\Lambda=0$ for simplicity.) In the AdS/CFT proposal, by contrast, the correspondence is between a gauge theory on the \emph{boundary} of an asymptotically anti-de Sitter space-time and a gravity theory in the \emph{bulk}. The gauge theory generically refers to a fixed  background metric (inherited from the AdS boundary), and the gravity theory has to have a negative cosmological constant.  Finally, because the proposal is \emph{much} more general --a large number of boundary gauge theories are considered and the bulk gravity theories vary accordingly-- in most cases one has good evidence supporting the correspondence, but not a detailed proof of equivalence as in our case.  

Our correspondence is perhaps closer in spirit to that in the  `double-copy' literature in which there is a detailed relation between perturbative scattering amplitudes of Yang-Mills theory and those of GR. As in our case, both theories refer to the bulk. Our correspondence emphasizes gauge invariance. The same is true in the double-copy correspondence because of its emphasis on the `on-shell' quantities. The two strategies seem complementary. One is non-perturbative and emphasizes Hamiltonian methods. The other is perturbative and emphasizes the S-matrix. Therefore, it is quite possible that the two will provide each other useful insights and hints as structures that naturally arise in one may be difficult to see in the other. For example, one could develop a perturbative expansion of the equations of motion in our approach --perhaps using an anti-self-dual background. Do they have any of the features seen in the double-copy literature?  

Our phase space is the same as in a Yang-Mills theory, and the Gauss and vector constraint parts of our Hamiltonian are the same as in the Yang-Mills theory (because, when appropriately written, they do not refer to a background metric). But the generators of `time evolution' is quite different in our background independent gauge theory from that in Yang-Mills. Ours is background independent, while the Yang-Mills Hamiltonian makes a crucial role of the background metric. Yet, perhaps a systematic perturbative expansion --paying due respect to all the coupling constants involved in the expansion--  will reveal that a relation naturally emerges.  In this respect the Lie algebra of GGC Lie derivatives (discussed in the Appendix) that dominate dynamics in our approach may be helpful to the double-copy investigations,  e.g., along the lines of  \cite{sdym,mason2}. If so, our formulation may shed light on the `origin' of the gauge-kinematics duality that plays a central role in the double copy literature, and could also suggest directions to extend the double copy results beyond perturbation theory.

For simplicity, in this paper we set various constants to unity.  On the other hand  the double copy results involve perturbative expansions. Therefore to facilitate the comparison it is useful to restore the dimensions. Denoting the gauge theory coupling constant by $\alpha$, setting $8\pi G =  \kappa$ and $c=1$ (but \emph{not} setting $\hbar=1$ since we are comparing classical theories here) we have the following relations:
\be \hbox{\rm Gauge Theory:} \quad \D_{a} \lambda_{i} = \partial_{a} \lambda_{i} + \alpha\, \ez_{ij}{}^{k}\, A_{a}^{j} \lambda_{k}, \quad{\rm and}\quad F_{ab} = 2\partial_{[a} A_{b]}^{i} + \alpha\, \ez_{ij}{}^{k}\, A_{a}^{j}A_{b}^{k} \ee
where the gauge theory connection $A_{a}^{i}$ has dimensions $M^{\f{1}{2}} L^{-\f{1}{2}}$ and the coupling constant $\alpha$ has dimensions  $(ML)^{-\f{1}{2}}$. On the GR side, we have:
\be \hbox{\rm Gravity:} \quad \D_{a} \lambda_{i} = \partial_{a} \lambda_{i} + \kappa \, \ez_{ij}{}^{k}\,\, {}^{G}\!\!A_{a}^{j} \lambda_{k}, \quad{\rm and}\quad  {}^{G}\!F_{ab} = 2\partial_{[a} ,\, {}^{G}\!\!A_{b]}^{i} + \kappa\, \ez_{ij}{}^{k}\,\, {}^{G}\!\!A_{a}^{j}\,\, {}^{G}\!\!A_{b}^{k} \ee
where now the gravity connection ${}^{G}\!\!A_{a}^{i}$ has dimensions $M L^{-2}$ and the coupling constant $\kappa$ has dimensions  $M^{-1} L$. Hence the relation between the connections and their conjugate momenta $\E^{a}_{k}$ are:
\be   {}^{G}\!\!A_{a}^{i}  =\f{\alpha}{G} A_{a}^{i} \qquad {\rm and} \qquad 
           {}^{G}\! \E^{a}_{i}=\f{G}{\alpha}  \E^{a}_{i}\, . \ee
Therefore after restoring the dimensionfull constants, the GR Hamiltonian, written as a function on $\psymc$, is given by
\be \label{ham4} H_{0,\Sv, \N} = \f{\kappa}{\alpha} \int_{\Sigma} \rmd^{3}x \, \big[ {\textstyle{\f{1}{2}}} \, \N\, \ez^{ij}{}_{k}\, \E^{a}_{i} \E^{b}_{j} \,F_{ab}{}^{k}\big] \, +\, \int_{\Sigma} \rmd^{3}x \,\, \big[S^{a} \E^{b}_{k}\, F_{ab}{}^{k} \big]  \nonumber\\
\ee
where the fields on the right side refer to gauge theory. Note that the coefficient in front of the crucial scalar constraint is $\f{\kappa}{\alpha}$; so strong coupling in the gauge theory corresponds to weak coupling on the GR side and vice versa. However, one has also to keep in mind, that the scalar part, that is responsible for `pure time evolution', refers to a background independent gauge theory and not to the Yang-Mills Hamiltonian. \smallskip

3. \emph{Implementation of the Dirac program in canonical gravity:} In LQG, quantum dynamics is pursued using two complementary approaches: The spinfoams that use a sum over histories approach \cite{perez-review,crfvbook}, and canonical quantum gravity that aims to complete the Dirac program for quantization of constrained systems. Both approaches use the fully developed  kinematic framework \cite{alm2t,alrev,ttbook,30years} which provides a Hilbert space of states and operators using background independent  tools from gauge theories. In the canonical approach, in particular, these  tools have provided concrete candidates for representing the constraints. The mathematical precision of these results is high in that technical issues related to infinite dimensional spaces have been addressed in detail; the constraints are not the formal, unregulated operators one finds, e.g., in the Wheeler-Dewitt theory. In particular, a careful analysis and the application of novel techniques has led to  background independent constructions of  the scalar constraint operator \cite{tt}.  However, the program has remained unsatisfactory in a crucial respect: there is a large freedom in the choice of this operator and so far none leads to a quantum constraint algebra that adequately mirrors the classical Poisson bracket algebra of two scalar constraints. 

Therefore, over the last decade, these methods have been further developed and applied to  progressively more complicated  models that capture more and more aspects of 3+1 GR \cite{pft1,pft2,hk,2+1u13,3+1u13,mvprop}. In these models classical dynamics can be understood in terms of geometrical  deformations of field variables in \emph{space-like} directions. Progress in the quantum theory of these models is then greatly facilitated by the incorporation of this property of classical evolution into the quantum dynamics driven by their Hamiltonian constraints.  The geometric insights into  the action of the scalar constraint of GR, reported in this paper, indicate that the key property of the classical evolution of model systems also holds for gravity. Results obtained in the model systems strongly suggest that the quantum scalar constraint should be constructed by appropriately incorporating this property. In the Riemannian case this has lead to the definition of a scalar constraint operator in which the operator analog of the electric shift introduced in this paper plays a central role. As a result, the action of the scalar constraint operator on quantum states is closely related to spatial diffeomorphisms. An immediate consequence of this geometrical implementation of the constraint action turns out to be the elimination of Perez's `spin ambiguity' \cite{perez}. Moreover this geometrical understanding of the constraint action points to the exciting possibility of a demonstration of a non-trivial anomaly free quantum constraint algebra  that mirrors the classical Poisson algebra.

In the Lorentzian case the geometrical understanding of classical evolution in terms of GGC Lie derivatives  with respect to the electric shift is again based on self-dual connections.  As noted in Section \ref{s2} and emphasized in \cite{sam}, these variables have a natural spacetime interpretation.  However, self-dual connections are now complex-valued while  the \emph{quantum} kinematics of LQG is based on real connections. So far it has proven difficult to generalize the underlying functional analysis to complex connections.  Consequently, a direct construction of a mathematically rigorous constraint operator in terms of the electric shift, similar to the Riemannian case,  seems out of reach at the moment. The question then is, if there is an indirect way to combine the rigor of the quantum framework based on the real variables  together with geometric insights on dynamics afforded by the complex variables. Remarkably, an affirmative answer exists to this question in the form of Thiemann's `complexifier' \cite{wick-tt}. The complexifier is a complex-valued function on the real phase space that generates a  canonical transformation from the real variables (used in the  Riemannian self-dual sector $\psymr$) to the complex variables (used in the Lorentzian self-dual sector $\psyml$).  Since the Riemannian and Lorentzian constraints arise simply as restrictions to the $\psymr$ and $\psyml$ sections of the constraints on the full complex phase space $\psymc$, they have the same \emph{form}. Hence, the canonical transformation generated by the complexifier can be thought of as a generalized Wick transform on {\em the gauge theory phase space}. The proposal is to implement this finite transformation in the quantum theory of the Riemannian sector and thereby directly map physical states in the kernel of  the quantum constraints of the Riemannian sector to physical states in kernel of the constraints in the Lorentzian sector \cite{wick-aa,wick-tt,wick-mv}. \medskip

These three remarks illustrate the implications and possible applications of our main results. There are also other potential applications of this gauge theory formulation of GR. For example, while in the space-time picture, discrete symmetries like parity and time reversal do not admit a natural extension to GR,  the gauge theory formulation makes it possible to introduce them because the operations are well defined in the internal space in the covariant version of this  framework. Similarly,  in geometrodynamics the scalar constraint has a `potential' term $\mathcal{R}$, in addition to the `kinetic' term that is quadratic in momenta $\P^{ab}$,  while in connection dynamics it is purely quadratic in momenta. Therefore, there is a precise sense in which dynamics of GR is encoded in the null geodesics of the (contravariant) supermetric $\ez_{ij}{}^{k}\,F_{ab}{}^{k}$ on the `connection superspace'  (even though it is degenerate at non-generic points of the phase space). These features could be used to gain new insights into the dynamics of GR. 

\section*{\bf Acknowledgements:} 
This work was supported by  the NSF grants PHY-1505411 and PHY-1806356 and the Eberly Chair funds of Penn State. MV would like to thank Fernando Barbero and  Eduardo Villase\~nor for discussions in the initial stage of this work.  AA is grateful to Alok Laddha for a number of discussions, both on LQG and the double copy program, and for his comments on the manuscript.

\begin{appendix}
\label{a1}

\section{Generalized gauge covariant Lie derivatives: Underlying structure}

In this Appendix we show, through explicit construction, that the action of generalized Gauge covariant (GGC)  Lie derivatives can be embedded into a graded Lie algebra structure, thus generalizing the Lie algebra structure defined by the action of ordinary Lie derivatives on tensor fields. Our considerations build on the exposition, viewpoint and treatment of gauge fields in \cite{ahm} (referred to as `AHM' in what follows).  Section \ref{secA1} provides a quick review of this framework. Section \ref{secA2}  recalls the Lie algebra structure of gauge covariant (GC) Lie derivatives with respect to ordinary vector fields. Section \ref{secA3} discusses the new element introduced in this paper: the graded Lie algebra structure for GGC Lie derivatives. 

\subsection{A short review of the AHM framework} 
\label{secA1}

Penrose and Geroch introduced an abstract index notation  and  coordinate-free, algebraic viewpoint   for ordinary tensor fields \cite{rp,rpwr,rg}. This framework was extended to include 
gauge fields that also carry internal indices  through the definition of {\em generalized} tensor fields $T^{a_1..a_A}{}_{b_1...b_B}{}^{i_1..i_P}{}_{j_1...j_Q}$.  Here, indices from the beginning of the alphabet denote the abstract indices that refer to ordinary tensors, while those from the middle of the alphabet represent internal indices that refer to a specified representation of a gauge group. The generalized tensor field $T^{a_1..a_A}{}_{b_1...b_B}{}^{i_1..i_P}{}_{j_1...j_Q}$ will be said to be of type $(A,B; P,Q)$;  its indices are place holders which indicate the type of tensor rather than component labels. Tensors of type $(0,0,0,0)$ are identified with functions $f$ on the manifold.

The action of a derivative operator on an $(A,B;P,Q)$ generalized tensor yields a  linear map to $(A,B+1; P, Q)$ generalized tensors subject to the Leibniz rule and the torsion free condition on functions. It follows that the action of a skew symmetric combination of derivative operators on $v_c{}^i$ is a functionally linear map since ${\cal D}_{[a }{\cal D}_{b]}f v_c{}^i = f {\cal D}_{[a }{\cal D}_{b]}v_c{}^i$. Therefore, it defines the internal and tensorial curvatures $F_{ab}{}^i{}_j$ and $R_{abc}{}^d$ through:
\be
2\,{\cal D}_{[a }{\cal D}_{b]}v_c{}^i = F_{ab}{}^i{}_jv_c{}^j + R_{abc}{}^dv_d{}^j
\label{curvf}
\ee
Let $\partial_a$ denote a derivative operator which is flat, i.e., such  that both $R_{abc}{}^d$ and $F_{ab}{}^i{}_j$ vanish. As mentioned in the main text, for simplicity we shall restrict attention to  derivative operators ${\cal D}_a$ whose actions on ordinary tensors  agrees with that of $\partial_a$. The difference  between the actions of ${\cal D}_a$ and $\partial_a $ on a field $v^i$ with an internal index defines a gauge connection $A_a{}^i{}_j$ through:
\be
{\cal D}_a v^i = \partial_a v^i + A_a{}^i{}_jv^j.
\label{conn}
\ee
In what follows we restrict attention to the gauge group $SU(2)$ and work in the adjoint (i.e. spin 1) representation. As in the main text, we restrict ourselves to derivative operators ${\cal D}, \partial$ which kill the Cartan-Killing metric on $su(2)$ (denoted by $\qz_{ij}$ in the main text). Therefore we can freely raise and lower internal indices also of fields that are `inside' derivatives, and the connection 1-form $A_a{}^i{}_j$ as well as the curvature 2-form $F_{ab}{}^{i}{}_{j}$ are skew symmetric in the internal indices.

\emph{Remark:} The key strength of the AHM framework is that  generalized tensors  are coordinate and gauge {\em invariant} objects whose  differential and algebraic manipulations can be performed without explicitly referring to bundles, trivializations, and sections. Behavior under changes of coordinates emerges from the choice of  a basis of vectors $\{e^{a}_{\mathbf{a}},\,\, {\mathbf{a} =1,2,3}\}$. Expansion of ordinary tensors in this basis and its dual yields components which change in the standard way under changes of coordinates. Similarly,  gauge  transformations emerge when an orthonormal basis of internal vectors $\{e^i_{\bf j}, {\bf j}=1,..,N\}$ and its dual are used to expand the `internal' part of generalized tensors.  (Orthonormality refers to the Cartan-Killing metric on the Lie algebra which is assumed to be semi-simple.) The coefficients in this expansion are components which change under the change of internal basis. A  trivialization in the language of bundles corresponds to a choice of basis.  As shown in \cite{ahm} such a basis defines a unique $\partial_a$ such that $\partial_a$ kills each element of the basis, i.e. $\partial_a e^i_{\bf j}=0,\, {\bf j}=1,..,N$.  The standard gauge transformation properties of components of various objects of interest then follow from the invariance of the corresponding generalized tensors in conjunction with change of basis and, for the connection, additionally, the change in the  corresponding $\partial_a$. 

\subsection{Lie algebra for GC Lie derivatives}
\label{secA2}

As discussed in Section \ref{s3.1}, the GC Lie derivative is defined by replacing the ordinary derivative $\partial_a$ by the gauge covariant derivative ${\cal D}_a$. The action of a GC Lie derivative on a generalized tensor field then yields a generalized tensor field of exactly the same index structure. Consider the action of the GC Lie derivative with respect to a vector field $v^a$ on the `internal' covector $\alpha_i$:
\be
{\gcl}_{\vec V} \alpha_i = V^a \D_a \alpha_i\, .
\label{gclaction}
\ee
It immediately follows that 
\ba
{\gcl}_{\vec U} {\gcl}_{\vec V}\alpha_i = U^a \D_a (V^b\D_b\alpha_i) &=& (U^a\partial_a V^b) \D_b\alpha_i+ U^aV^b\D_a\D_b \alpha_i
\nonumber\\
{\rm whence,}\quad  [{\gcl}_{\vec U}, {\gcl}_{\vec V}]\alpha_i 
&=& [U, V]^a \D_a \alpha_i +   U^aV^b F_{ab\,i}{}^j \alpha_j \, .
\label{comgcl}
\ea
Whereas the first term is a gauge covariant Lie derivative with respect to the (usual) commutator of vector fields, the second term in (\ref{comgcl}) is {\em not} a gauge covariant Lie derivative. Hence  the operation $\gcl$ is not closed under commutator. Note however that since $F$ takes values in $su(2)$, the second term corresponds to the action of the infinitesimal gauge transformation $U^aV^b F_{abi}{}^j$ on $\alpha_j$. This suggests the possibility of embedding the action (\ref{gclaction}) into a larger structure that then constitutes a Lie algebra. Define the linear operator ${\hat O}_{(V, \Phi)}$ with a vector field $V^a$  and an $su(2)$ gauge transformation $\Phi_i{}^j$  which acts on $\alpha_i$ as:
\be
{\hat O}_{({\vec V}, \Phi)}  \alpha_i = {\gcl}_{\vec V} \alpha_i + \Phi_i{}^j \alpha_j
\label{ogcl}
\ee
Note that just as the commutator of a pair of $su(2)$ gauge transformations is another $su(2)$ gauge transformation, the commutator of a gauge covariant derivative and an $su(2)$ transformation also  yields an $su(2)$ transformation:
\ba [\gcl_{\vec W},\, \Phi_i{}^j ]\, \alpha_j  &=&  W^a\D_a(\Phi_i{}^j\alpha_j)- \Phi_i{}^j\,W^a\D_a(\alpha_j)\, ,\quad {\rm whence,}  \nonumber \\
\hskip2cm [\gcl_{\vec W}, \Phi_i{}^j ]  &=&  (W^a\D_a \Phi_i{}^j) \equiv \gcl_{\vec W} \Phi_i{}^j\, .
\label{gcllcomm}
\ea
It is then straightforward to see that the commutator of any two operators ${\hat O}_{({\vec U}, \Theta)}$ and  ${\hat O}_{({\vec V}, \Phi)}$ is given by
\be
[{\hat O}_{({\vec U}, \Theta)} \;,\;{\hat O}_{({\vec V}, \Phi)}]  \alpha_i =  {\hat O}_{(\vec{W}, \Psi)} \alpha_{i}
\label{ogclcomm}
\ee
where 
\be 
W^{a} = [U,\,V]^{a}\quad {\rm and} \quad \Psi_{i}{}^{j} = \gcl_{\vec U}\Phi_{i}{}^{j} - \gcl_{\vec V}\Theta_{i}{}^{j}  + [\Theta, \Phi ]_i{}^j +U^aV^b F_{ab\,i}{}^j \, .
\label{ogcllie}
\ee
Thus the vector space generated by the linear operators ${\hat O}_{({\vec V_1}, \Phi_1)}$ is closed under the operation of taking commutators. Since these operators generate an associative algebra, it follows immediately that the commutator bracket satisfies the Jacobi identity:
\be 
[\;{({\vec V_1}, \Phi_1)},\; [{({\vec V_2}, \Phi_2)},{({\vec V_3}, \Phi_3)}]\;] + \hbox{\rm cyclic permutations} =0\,
\ee
One can also check the Jacobi identity by explicitly evaluating the double commutators and using the Bianchi identity satisfied by $F_{ab}{}^{i}{}_{j}$. Thus the infinite dimensional vector space of pairs of spatial vector fields  and infinitesimal gauge transformations is endowed with a Lie bracket.

\subsection{\label{secA3} Lie algebra for GGC Lie derivatives}

Recall from Section \ref{s3.2} that the GGC Lie derivative with respect to a generalized vector field $V^a{}_{i_1..i_R}$ of a generalized tensor of type $(M,N,P,Q)$ is defined by:\\
\noindent (i) Writing down the expression for the ordinary Lie derivative with respect to a vector field $V^a$ of an ordinary tensor of type $(M,N)$, and\\
\noindent (ii) Replacing every instance of an ordinary derivative in (i) by the gauge covariant derivative $\D$, every instance of $V^a$ by 
$V^a{}_{i_1..i_R}$ and every instance of the $(M,N)$ tensor by the $(M,N,P,Q)$ tensor.\\
From (i) and (ii), the action of the GGC Lie derivative with respect to a vector field $V^a_k$ on the `internal' covector $\alpha_i$ is
\be
{\ggcl}_{\vec V_k} \alpha_i = V^a_k \D_a \alpha_i\, .
\label{ggclaction}
\ee
Note that the GGC Lie derivative {\em no longer} maps a generalized tensor field to one of the same index structure. Instead it {\em adds} an extra lower internal index. A second such GGC Lie derivative would add yet another lower internal index. Therefore to obtain a closed commutator, the we need to consider GGC Lie derivatives generated by vector fields carrying an arbitrary number of lower indices. The fact that any putative Lie bracket between a pair of generalized vector fields with $m$ and $n$ indices must result in a generalized vector field with $m+n$ internal indices, implies that the resulting Lie algebra is {\em graded} with the grading corresponding to the number of internal indices of these generalized vector fields. Since the Jacobi identity involves {\em double} commutators, the resulting proliferation of indices makes a direct brute force check an extremely involved exercise. Although we have directly checked the Jacobi identity for the commutator bracket (\ref{liebrktfinal}) given below, for brevity,  we will establish it simply by showing that our Lie bracket is the commutator bracket of linear operators in an associative algebra, and the vector space of these operators is closed under the commutator.

Let us calculate the commutator of two GGC Lie derivatives. Since
\be
{\ggcl}_{\vec V_{k_1k_2..k_M}} \alpha_{i_1..i_Q} = V^a_{k_1k_2..k_M} \D_a \alpha_{i_1..i_Q} ,
\label{ggclactionmq} 
\ee
we have
\ba
 {\ggcl}_{\vec U_{l_1l_2..l_N}} {\ggcl}_{\vec V_{k_1k_2..k_M}}\alpha_{i_1..i_Q} &=& U^a_{l_1l_2..l_N} \D_a (V^b_{k_1k_2..k_M}\D_b\alpha_{i_1..i_Q}) \nonumber\\
&=& (U^a_{l_1l_2..l_N}\D_a V^b_{k_1k_2..k_M}) \D_b\alpha_{i_1..i_Q}
+ U^a_{l_1l_2..l_N}
V^b_{k_1k_2..k_M}\D_a\D_b \alpha_{i_1..i_Q} \, .
\nonumber
\ea
Therefore, the commutator is given by
\ba
[{\ggcl}_{\vec U_{l_{1},l_{2} \ldots l_{N}}}, \, {\ggcl}_{\vec V_{k_1k_2 \ldots k_M}} ] \alpha_{i_1..i_Q}
&=& [U_{l_1l_2 \ldots l_N}, \,\,V_{k_1k_2..k_M}]^a \D_a \alpha_{i_1\ldots i_Q} \nonumber \\
&+& U^a_{l_1l_2\ldots l_N} V^b_{k_1k_2..k_M} 
\sum_{I=1}^Q F_{abi_{I}}{}^{j_I}\alpha_{i_1\ldots i_{I-1}j_Ii_{I+1}....i_Q}\;\;\;\;\;\;\;\;\;\;\;\;
\label{comggcl}
\ea
where we have defined the `gauge covariant' commutator of a pair of generalized vector fields in the obvious way:
\be
[U_{l_1l_2..l_N},\, V_{k_1k_2..k_M}]^a := U^b_{l_1l_2..l_N}\D_b V^a_{k_1k_2..k_M} - V^b_{k_1k_2..k_M}\D_b U^a_{l_1l_2..l_N} \equiv  \ggcl_{\vec U_{l_1l_2..l_N}} V^a_{k_1k_2..k_M}\, .
\label{defcomuvnm}
\ee
As in the case of GC Lie derivatives now  the first term of (\ref{comggcl}) is a GGC Lie derivative, and the second 
term is an infinitesimal gauge rotation of $\alpha$. Note however, that in contrast to the case of GC Lie derivatives (\ref{comgcl}), the gauge rotation is itself a generalized tensor with $N+M$ internal indices. This suggests that we define a linear operator on $\alpha$ which augments the GGC Lie derivative action with the appropriate $su(2)$ rotation by a generalized tensor. Accordingly let $\Phi_{(M)}$  denote such a generalized tensor valued gauge rotation so that 
$\Phi_{(M)} \equiv \Phi_{k_1k_2..k_M}{}_{i}{}^{j}$  that is anti-symmetric 
in its ${}_i{}^j$ `rotator' indices. Define, in obvious notation, the action of the linear operator ${\hat O}_{({\vec V}_{(M)}, \Phi_{(M)})}$ as:
\be
{\hat O}_{({\vec V}_{(M)}, \Phi_{(M)})} \alpha_{i_1..i_Q} = 
{\ggcl}_{\vec V_{k_1k_2..k_M}} \alpha_{i_1..i_Q} + \sum_{I=1}^Q\Phi_{k_1k_2..k_M}{}_{i_I}{}^{j_I}\alpha_{i_1...i_{I-1}j_Ii_{I+1}....i_Q} .
\label{ggclgmq}
\ee
In order to cut down on algebraic clutter in the computation of the commutator of two such operators we define the following notation
\ba
\Phi_{(M)} \cdot \alpha_{(Q)} & \equiv &\sum_{I=1}^Q\Phi_{k_1k_2..k_M}{}_{i_I}{}^{j_I}\alpha_{i_1...i_{I-1}j_Ii_{I+1}....i_Q} \nonumber\\
&\equiv &  \sum_{I=1}^Q\Phi_{(M)}{}_{i_I}{}^{j_I}\alpha_{i_1...i_{I-1}j_Ii_{I+1}....i_Q}\quad {\rm and}  \label{notation1} \\
\ggcl_{\vec V_{(M)}} \alpha_{(Q)} &\equiv &{\ggcl}_{\vec V_{k_1k_2..k_M}} \alpha_{i_1..i_Q} 
\label{notation2}
\ea
Using this notation, we have the following useful Lemma:\medskip

\noindent{\bf Lemma}: 
\be
\ggcl_{{\vec V}_{(M)}} (\Phi_{(N)} \cdot \alpha_{(Q)}) = 
\Phi_{(N)}\cdot (\ggcl_{{\vec V}_{(M)}}\alpha_{(Q)}) \,-\, \ggcl_{ \Phi_{(N)}\cdot{\vec V}_{(M)} }\,\,\alpha_{(Q)} \,+\, (\ggcl_{{\vec V}_{(M)}} \Phi_{(N)})\cdot \alpha_{(Q)}
\label{l1}
\ee
where in the second term above, and here on,  it is understood that $\Phi_{(N)}$ ignores the tangent space indices 
(here of $V^{a}_{(M)}$)  and acts only on internal indices as in (\ref{notation1}).
\medskip

\noindent{\bf Proof}: It is straightforward to see that:
\be
\ggcl_{{\vec V}_{(M)}} (\Phi_{(N)} \cdot \alpha_{(Q)})  = {V}_{(M)}^a\D_a (\Phi_{(N)} \cdot \alpha_{(Q)}) = ({V}_{(M)}^a\D_a \Phi_{(N)}) \cdot\alpha_{(Q)}
+ {V}_{(M)}^a (\Phi_{(N)}\cdot (\D_a \alpha_{(Q)}))
\label{l1.0}
\ee
The first term can be written as a GGC Lie derivative:
\be
({V}_{(M)}^a\D_a \Phi_{(N)}) \cdot \alpha_{(Q)} =: (\ggcl_{{\vec V}_{(M)}} \Phi_{(N)})\cdot \alpha_{(Q)}
\label{l1.1}
\ee
Since the GGC Lie derivative kills the Cartan-Killing metric, it follows that $\ggcl_{{\vec V}_{(M)}} \Phi_{(N)}{}_i{}^j$ is anti-symmetric on its  ${}_i{}^j$ indices and hence corresponds to  a generalized tensor valued $su(2)$ rotation. 
The second term can be evaluated as follows.
\ba
{V}_{(M)}^a \Phi_{(N)}\cdot (\D_a \alpha_{(Q)}) &=& {V}_{(M)}^a   \sum_{I=1}^Q\Phi_{(N)}{}_{i_I}{}^{j_I}\D_a\alpha_{i_1...i_{I-1}j_Ii_{I+1}....i_Q} \nonumber\\
&=&    \sum_{I=1}^Q\Phi_{(N)}{}_{i_I}{}^{j}{V}_{(M)}^a\D_a\alpha_{i_1...i_{I-1}ji_{I+1}....i_Q} + \sum_{I=1}^M\Phi_{(N)}{}_{k_I}{}^{j_I} V^a_{k_1...k_{I-1}j_Ik_{I+1}....k_M}
\D_a\alpha_{(Q)}  \nonumber \\
&-&
\sum_{I=1}^M\Phi_{(N)}{}_{k_I}{}^{j_I} V^a_{k_1...k_{I-1}j_Ik_{I+1}....k_M}
\D_a\alpha_{(Q)}  \nonumber \\
&=& \Phi_{(N)}\cdot (\ggcl_{{\vec V}_{(M)}}\alpha_{(Q)}) - \ggcl_{ \Phi_{(N)}\cdot{\vec V}_{(M)} }\alpha_{(Q)}\;\;\;\;\;\;\;\;\;\; \label{l1.2}
\ea
Equation (\ref{l1}) follows from (\ref{l1.0}) -(\ref{l1.2}) 
\\

Finally, it is straightforward to verify that \
\be
\Theta_{(M)} \cdot ( \Phi_{(N)} \cdot \alpha_{(Q)}) - 
\Phi_{(N)} \cdot ( \Theta_{(M)} \cdot \alpha_{(Q)})
= \big(\Theta_{(M)} \circ \Phi_{(N)} -\Phi_{(N)} \circ \Theta_{(M)}\big) \cdot \alpha_{(Q)}
\label{compos1}
\ee
where we have defined:
\ba
&&\Big(\Theta_{(M)} \circ \Phi_{(N)} -\Phi_{(N)} \circ \Theta_{(M)}\Big)_{i}{}^j :=  \sum_{J=1}^N  \Theta_{(M)}{}_{n_J}{}^{{\bar n}^J} \Phi_{n_1..n_{J-1}{\bar n}_Jn_{J+1}..N}{}_{\,i}{}^j\nonumber\\
&&- \sum_{K=1}^M \Phi_{(N)}{}_{m_K}{}^{{\bar m}^K} \Theta_{m_1..m_{K-1}{\bar m}_Km_{K+1}..M}{}_{\,i}{}^j
+ \Theta_{(M)}{}_i{}^k \Phi_{(N)}{}_k{}^j - \Phi_{(N)}{}_i{}^k \Theta_{(M)}{}_k{}^j .
\label{compos2}
\ea
Note that the right hand side of the above equation is anti-symmetric in its ${}_i{}^j$ `rotator' indices.

We are now ready to evaluate the commutator of 2 linear operators of the type (\ref{ggclgmq}). First we note that 
\ba
{\hat O}_{({\vec U}_{(M)}, \Theta_{(M)})} {\hat O}_{({\vec V}_{(N)}, \Phi_{(N)})}  \alpha_{(Q)} &=&
\ggcl_{ {\vec U}_{(M)}} (\ggcl_{ {\vec V}_{(N)} } \alpha_{(Q)} 
+  \Phi_{(N)}\cdot  \alpha_{(Q)}  ) \nonumber\\
&+ &
    \Theta_{(M)}\cdot  (\ggcl_{{\vec V}_{(N)}}  \alpha_{(Q)} )  + 
\Theta_{(M)} \cdot ( \Phi_{(N)} \cdot \alpha_{(Q)}) \, .
\;\;\;\;\;\;\;\;\;\;
\ea
Therefore the commutator is given by:
\ba
[{\hat O}_{({\vec U}_{(M)}, \Theta_{(M)})},\,\, {\hat O}_{({\vec V}_{(N)}, \Phi_{(N)})} ] \alpha_{(Q)}
&=& [\ggcl_{ {\vec U}_{(M)}}, \ggcl_{ {\vec V}_{(N)} }]  \alpha_{(Q)}  
+  \ggcl_{ {\vec U}_{(M)}}  (\Phi_{(N)}\cdot  \alpha_{(Q)}) \nonumber \\
&-& \ggcl_{ {\vec V}_{(N)}}  (\Theta_{(M)}\cdot  \alpha_{(Q)} ) +
\Theta_{(M)}\cdot  (\ggcl_{{\vec V}_{(N)}}  \alpha_{(Q)} )  - \Phi_{(N)}\cdot  (\ggcl_{{\vec U}_{(M)}}  \alpha_{(Q)} ) \nonumber \\
 &+& \Theta_{(M)} \cdot ( \Phi_{(N)} \cdot \alpha_{(Q)}) - 
\Phi_{(N)} \cdot ( \Theta_{(M)} \cdot \alpha_{(Q)})\, .
\ea
Using (\ref{comggcl})  and the Lemma (\ref{l1}) in the first line and (\ref{compos1}) in the second, it is straightforward to see that the expression of the  commutator reduces to
\ba
[{\hat O}_{({\vec U}_{(M)}, \Theta_{(M)})} {\hat O}_{({\vec V}_{(N)}, \Phi_{(N)})} ] \alpha_{(Q)} &=&
\ggcl_{ [{\vec U}_{(M)},  {\vec V}_{(N)}] }  \alpha_{(Q)}  + F(U_{(M)}V_{(N)}) \cdot \alpha_{(Q)}
\nonumber \\
%&&+   \ggcl_{ {\vec U}_{(M)}}  \Phi_{(N)}\cdot  \alpha_{(Q)}    - \ggcl_{ {\vec V}_{(N)}}  \Theta_{(M)}\cdot  \alpha_{(Q)} 
%\nonumber \\
&+& (\ggcl_{\Theta_{(M)}\cdot{\vec V}_{(N)}}     - \ggcl_{\Phi_{(N)}\cdot{\vec U}_{(M)}})  \alpha_{(Q)}
 +  ( \ggcl_{{\vec U}_{(M)}}\Phi_{(N)}
-\ggcl_{{\vec V}_{(N)}}\Theta_{(M)} )\cdot \alpha_{(Q)}\nonumber \\
&+& 
(\Theta_{(M)} \circ \Phi_{(N)} -\Phi_{(N)} \circ \Theta_{(M)}) 
\cdot \alpha_{(Q)}\;\;\;\;\;\;\;\;\;\;
\label{lieggcl}
\ea
where we have defined 
\be
F( U_{(M)}V_{(N)})_i{}^j \, := \, U^a_{(M)} V^b_{(N)} F_{abi}{}^j .
\ee
Since the right side of (\ref{lieggcl}) is again a linear combination of GGC Lie derivatives and gauge rotations, the vector space generated by the linear operators  ${\hat O}_{({\vec U}_{(M)}, \Theta_{(M)})}$ is closed under the commutator bracket.  Let us make the closure explicit by rewriting the commutator as
\be \label{liebrktfinal}  [ ({\vec U}_{(M)}, \,\,\Theta_{(M)}),({\vec V}_{(N)}, \Phi_{(N)})] = ({\vec W}_{(M+N)}, \,\,\Psi_{(M+N)}) \ee
where 
\ba {\vec W}_{(M+N)} &=& { [{\vec U}_{(M)},  {\vec V}_{(N)}] } + {\Theta_{(M)}\cdot{\vec V}_{(N)}} - {\Phi_{(N)}\cdot{\vec U}_{(M)}} \quad {\rm and} \nonumber\\
\Psi_{(M+N)} &=&  F(U_{(M)}V_{(N)}) + \ggcl_{{\vec U}_{(M)}}\Phi_{(N)}
-\ggcl_{{\vec V}_{(N)}}\Theta_{(M)} +
\Theta_{(M)} \circ \Phi_{(N)} -\Phi_{(N)} \circ \Theta_{(M)} 
 \nonumber\ea
(\ref{liebrktfinal}) is a Lie bracket since the satisfaction of the Jacobi Identity is an immediate consequence of the associativity of the product of linear operators ${\hat O}_{({\vec U}_{(M)}, \Theta_{(M)})}$ on the vector space of fields with only internal indices. %We have also confirmed the Jacobi identity through a direct brute force calculation.

Let us summarize the main result of this Appendix.  Consider the vector space ${\cal V}_n$  of generalized vector fields with a fixed number $n$ of internal indices so that ${\cal V}_n=\{v^a_{i_1..i_n}\}$. The space of generalized vector fields with an arbitrary number of indices ${\cal V}= \bigoplus_{n=1}^{\infty} {\cal V}_n$ is a graded vector space with grading measured by $n$. Similarly, we have a graded vector space $\cal{I}$ of internal rotations $\Theta_{k_{1} \ldots k_{N}}{}_{\,i}{}^{j}$, with arbitrary $N$. In this section we have shown that  the bracket (\ref{liebrktfinal}) defines a graded Lie algebra on ${\cal V} \oplus {\cal I}$.

Our analysis focussed on  the  commutator action  (\ref{lieggcl})  on generalized tensors with only (an arbitrary number of)  covariant internal indices. Since the  Cartan-Killing Metric is killed by $\D$ and is invariant under $SU(2)$ gauge transformations, it follows that the analysis also holds for arbitrary generalized tensors with an arbitrary number of covariant and contravariant internal indices.  It would be interesting to investigate whether our main result can be extended to the action of GGC Lie derivatives on generalized tensors that also have ordinary indices.  Can this action also be embedded into a larger Lie algebra, e.g.,  constructed from commutators of appropriate operators as in this Appendix, is an interesting open issue in mathematical physics.

\end{appendix}

\end{document}